%% file: tmi.tex
\documentclass[journal,twoside,web]{ieeecolor}

\usepackage{cite}
\usepackage{amsmath,amssymb,amsfonts}
\usepackage{algorithmic}
\usepackage{graphicx}
\usepackage{textcomp}
\usepackage{orcidlink}

\usepackage{algorithm}
\usepackage{algorithmic}
\usepackage{multicol}
\usepackage{multirow}
\usepackage{graphicx}
\usepackage{amsmath}
\usepackage{amssymb}
\usepackage{pifont}
\newcommand{\cmark}{\ding{51}}
\newcommand{\xmark}{\ding{55}}

\usepackage{color}
\usepackage{xcolor}
\usepackage{colortbl}
\usepackage{nicematrix}

\usepackage{hyperref}
\usepackage{url}
\usepackage{graphicx}
\usepackage{wrapfig}
\usepackage{subcaption} 
\usepackage{multirow}
\usepackage{booktabs}

\definecolor{subsectioncolor}{rgb}{0,0.541,0.855}
\definecolor{nocolor}{rgb}{1,1,1}

\newcommand{\revise}[1]{\textcolor{black}{#1}}

\input{math_commands.tex}

\def\BibTeX{{\rm B\kern-.05em{\sc i\kern-.025em b}\kern-.08em
    T\kern-.1667em\lower.7ex\hbox{E}\kern-.125emX}}

\begin{document}

\title{
FairFedMed: Benchmarking Group Fairness in Federated Medical Imaging with FairLoRA 
}

\author{Minghan Li\orcidlink{0000-0003-0055-9261}, Congcong Wen\orcidlink{0000-0001-6448-003X}, Yu Tian\orcidlink{0000-0001-5533-7506}, Min Shi\orcidlink{0000-0002-7200-1702}, Yan Luo\orcidlink{0000-0001-5135-0316}, Hao Huang\orcidlink{0000-0002-9131-5854}, Yi Fang\orcidlink{0000-0001-9427-3883}, Mengyu Wang\orcidlink{0000-0002-7188-7126}
\thanks{This work was supported by NIH R01 EY036222 and NIH R21 EY035298.}
\thanks{
Minghan Li, Yan Luo, and Mengyu Wang are with Harvard AI and Robotics Lab and Harvard Ophthalmology AI lab, Harvard University, Boston, Massachusetts, USA (e-mail: mili4@meei.harvard.edu; yluo16@meei.harvard.edu; mengyu\_wang@meei.harvard.edu).
Congcong Wen is with the Harvard AI and Robotics Lab, Harvard University, Boston, MA, USA; the Embodied AI and Robotics Lab, New York University, New York, NY, USA; and the NYUAD Center for Artificial Intelligence and Robotics, New York University Abu Dhabi, Abu Dhabi, UAE.(e-mail: wencc@nyu.edu)
Hao Huang and Yi Fang are with the Embodied AI and Robotics Lab, New York University, New York, USA and NYUAD Center for Artificial Intelligence and Robotics, New York University Abu Dhabi, Abu Dhabi, UAE. (e-mail: hh1811@nyu.edu; yfang@nyu.edu).
Yu Tian is with the Department of Computer Science, University of Central Florida, Florida, USA (e-mail: yu.tian2@ucf.edu).
Min Shi is with the School of Computing and Informatics, University of Louisiana at Lafayette, Lafayette, LA 70504 USA (e-mail: min.shi@louisiana.edu)
}
\thanks{Minghan Li, Congcong Wen and Yu Tian are co-first authors.
}
}

\maketitle

\input{secs/0_abstract}


\input{secs/1_introduction}
\input{secs/2_related_work}
\input{secs/3_data_setup}  
\input{secs/4_method}

\input{secs/5_experiments}
\input{secs/6_conclusion}

{
\bibliographystyle{IEEEtran}
\bibliography{tmi}
}

\end{document}

%% file: math_commands.tex
\usepackage{amsmath,amsfonts,bm}

\def\eqref#1{equation~\ref{#1}}

\def\1{\bm{1}}

\def\mS{{\bm{S}}}

\def\mU{{\bm{U}}}
\def\mV{{\bm{V}}}
\def\mW{{\bm{W}}}

\DeclareMathAlphabet{\mathsfit}{\encodingdefault}{\sfdefault}{m}{sl}
\SetMathAlphabet{\mathsfit}{bold}{\encodingdefault}{\sfdefault}{bx}{n}



%% file: secs/0_abstract.tex
\begin{abstract}
Fairness remains a critical concern in healthcare, where unequal access to services and treatment outcomes can adversely affect patient health. While Federated Learning (FL) presents a collaborative and privacy-preserving approach to model training, ensuring fairness is challenging due to heterogeneous data across institutions, and current research primarily addresses non-medical applications. \revise{ To fill this gap, we establish the first experimental benchmark for fairness in medical FL, evaluating six representative FL methods across diverse demographic attributes and imaging modalities. We introduce \textit{FairFedMed}, the first medical FL dataset specifically designed to study group fairness (i.e., demographics). It comprises two parts: FairFedMed-Oph, featuring 2D fundus and 3D OCT ophthalmology samples with six demographic attributes; and FairFedMed-Chest, which simulates real cross-institutional FL using subsets of CheXpert and MIMIC-CXR. Together, they support both simulated and real-world FL across diverse medical modalities and demographic groups. Existing FL models often underperform on medical images and overlook fairness across demographic groups. To address this, we propose \textit{FairLoRA}, a fairness-aware FL framework based on SVD-based low-rank approximation. It customizes singular value matrices per demographic group while sharing singular vectors, ensuring both fairness and efficiency. }
Experimental results on the \textit{FairFedMed} dataset demonstrate that \textit{FairLoRA} not only achieves state-of-the-art performance in medical image classification but also significantly improves fairness across diverse populations. Our code and dataset can be accessible via the link: \url{https://wang.hms.harvard.edu/fairfedmed/}.
\end{abstract}

%% file: secs/1_introduction.tex
\section{Introduction}

\begin{figure*}[t]
    \centering
    \begin{subfigure}{0.395\textwidth}
        \centering
        \includegraphics[width=0.95\linewidth]{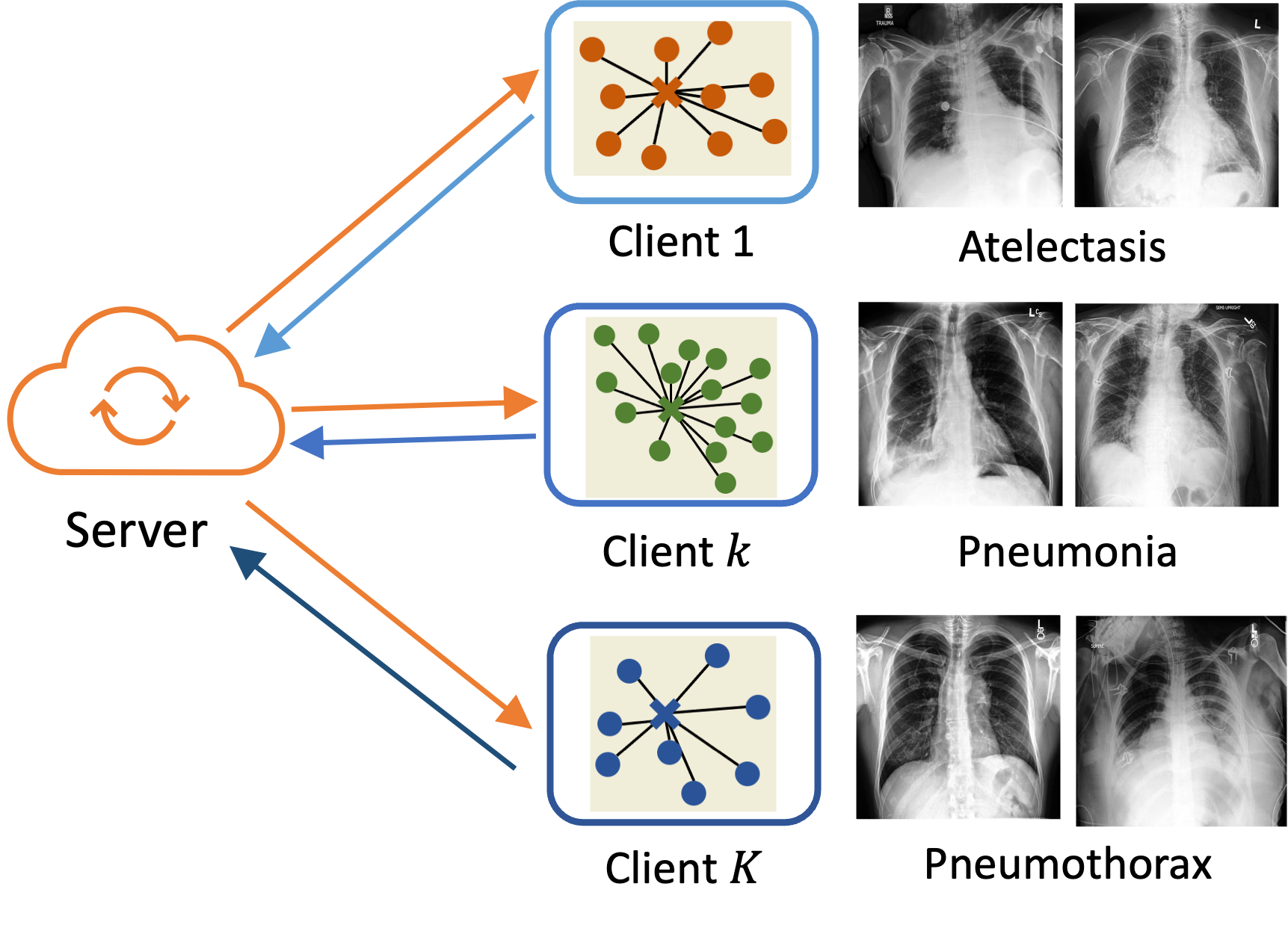}
        \vspace{-2mm}
        \caption{\revise{Site fairness in federated learning}}
    \end{subfigure}%
    \hfill
    \begin{subfigure}{0.57\textwidth}
        \centering
        \includegraphics[width=0.95\linewidth]{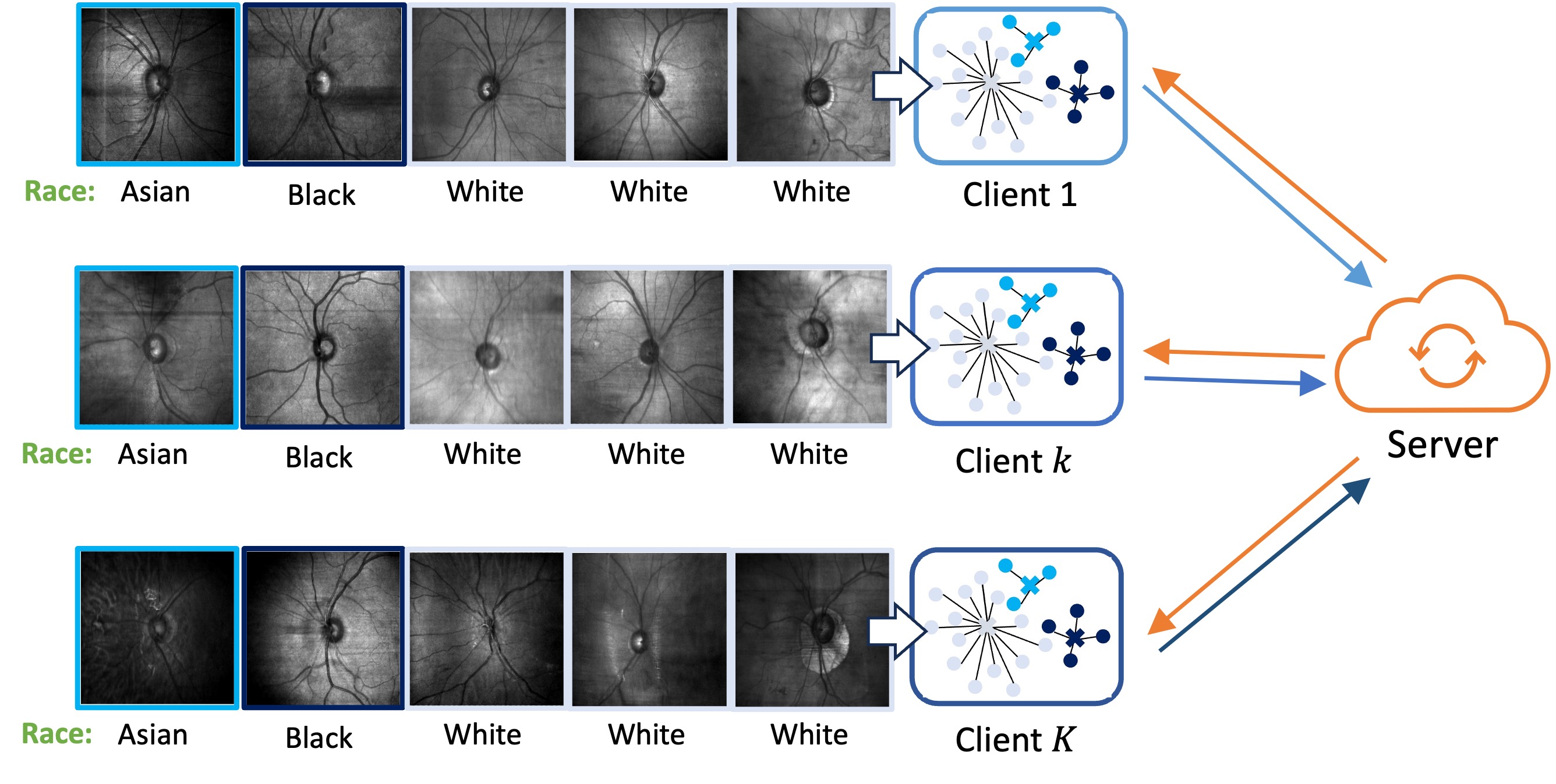}
        \vspace{-2mm}
        \caption{Group fairness in federated learning}
    \end{subfigure}%
    \vspace{-1mm}
    \caption{\revise{Comparison between site and group fairness in federated learning. Site fairness guarantees consistent model accuracy across each client's local data, such as different categories, while group fairness ensures equitable performance across all demographic groups, such as Asian, Black and White in the race attribute.
    }}
    \label{fig:site_group}
\end{figure*}

\revise{ Achieving group fairness in the medical domain is a highly challenging task, particularly in real-world scenarios where data distributions are significantly heterogeneous. Group fairness aims to ensure that models perform consistently across different demographic groups (e.g., race, gender, language), thereby avoiding algorithmic bias that could lead to misdiagnosis or underdiagnosis in certain populations—a risk with potentially serious clinical consequences. This issue is especially sensitive and critical in healthcare applications.
}

\revise{ 
Federated Learning (FL)~\cite{li2020federated_survey,lin2025fedrsclip} has emerged as a privacy-preserving distributed learning paradigm that enables collaborative model training across institutions without the need to share raw patient data. Unlike traditional centralized approaches, FL addresses the issue of data silos caused by privacy regulations and intellectual property concerns. This not only reduces the risk of data breaches but also enhances model generalizability by incorporating diverse data distributions from multiple sources. More importantly, FL has the structural potential to promote group fairness. By allowing model updates to be informed by decentralized yet demographically diverse data, FL provides a more balanced foundation for modeling across population groups. 
}

\revise{ 
Current fairness research in FL can be broadly categorized into two types: site fairness~\cite{lin2022personalized,cheng2024fedgcr}, which ensures consistent performance across participating institutions, and group fairness~\cite{agrawal2024no,badar2024fairtrade}, which focuses on equitable outcomes across demographic groups. 
Specifically, group fairness is broadly defined as the requirement that certain statistical properties of a model—such as accuracy, recall, or false positive rate—are similar or consistent across different demographic groups. In other words, the model should deliver balanced performance across groups to avoid unfair bias toward any particular population subgroup.
In healthcare, group fairness is particularly crucial, as patient distributions are inherently imbalanced in the real world. Neglecting minority populations in model development may result in clinical risks and raise serious ethical concerns.
Therefore, advancing group fairness in FL is not merely a technical challenge, it is a fundamental requirement for equitable healthcare and a matter of social responsibility.
}

Unfortunately, current research on group fairness in FL mainly focuses on non-medical tasks \cite{agrawal2024no,badar2024fairtrade}. 
In the medical field, cross-institution demographic differences are common due to geographic heterogeneity (e.g., hospitals in Asia and hospitals in North America), but research on how to manage these differences within the FL framework is relatively limited. \revise{So far, there is a lack of a comprehensive benchmark for group fairness in medical FL. Furthermore, although publicly available medical fairness datasets~\cite{irvin2019chexpert,johnson2019mimic,groh2021evaluating}, there is no dataset specifically designed for studying fairness in medical FL. Bridging the gap is key to advancing health equity.}

In this study, we present the first benchmark for evaluating fairness in medical federated learning, assessing six representative methods across diverse demographic attributes and imaging modalities. Furthermore, we introduce FairFedMed, the first comprehensive dataset specifically designed for studying fairness in medical FL. FairFedMed is composed of two components: FairFedMed-Oph, focused on ophthalmology, and FairFedMed-Chest, centered on chest X-ray imaging. These datasets are designed to support both simulated and real-world FL scenarios, spanning diverse medical modalities and rich demographic attributes to enable rigorous group fairness evaluation. FairFedMed-Oph contains \revise{ 16,681} real-world patient samples, each with paired 2D fundus (SLO) and 3D OCT images. It is the first fairness-aware FL dataset to include both 2D and 3D ophthalmic modalities. Each sample is annotated with six demographic attributes, and the dataset is partitioned into clients with diverse demographic distributions to support group fairness evaluation. FairFedMed-Chest simulates a real-world cross-institutional FL setting using chest X-ray subsets of CheXpert~\cite{irvin2019chexpert} and MIMIC-CXR~\cite{johnson2019mimic}, each as a separate client. Demographic shifts in race, age, and gender offer realistic evaluation scenarios. Together, these datasets form a strong foundation for fairness-aware FL research in medical imaging tasks.

While existing FL methods perform well on natural images, they often struggle with medical images due to domain-specific challenges\revise{\cite{pfitzner2021federated}}. Moreover, they lack mechanisms to ensure group fairness across diverse demographic groups. To overcome these limitations, we propose \textit{FairLoRA}, a fairness-aware FL framework for disease classification. This framework is based on CLIP \cite{radford2021CLIP} and introduces fairness into the low-rank approximation (LoRA) method. Specifically, to preserve unique intra-group characteristics, FairLoRA customizes singular value matrices for each demographic group while sharing singular vector matrices across all groups to capture inter-group relationships. Local models train a set of low-rank matrices on their local data, and the global model integrates matrices from different clients, allowing the model to aggregate global knowledge about demographic attributes from distributed data. This collaborative approach ensures that the model is not biased toward any particular client or demographic group. Experimental results on the FairFedMed dataset demonstrate that our model achieves state-of-the-art performance in medical image classification while ensuring equitable outcomes across demographic groups. 

Our main contributions are summarized as follows:
\begin{itemize}
\item \revise{ We establish the first experimental benchmark for fairness-aware medical FL by evaluating six representative FL methods on FairFedMed across diverse demographic attributes and imaging modalities.}

\item  \revise{ We introduce FairFedMed, the first medical FL dataset for group fairness that includes real-world ophthalmology data (2D fundus and 3D OCT) for simulated FL, and chest X-ray data for real cross-site FL—enabling comprehensive and practical fairness evaluations.}

\item  We propose FairLoRA, a group fairness-aware FL framework for disease classification. It customizes singular value matrices for each group to preserve unique characteristics while sharing singular vector matrices across groups to capture inter-group relationships.

\end{itemize}

%% file: secs/2_related_work.tex
\section{Related Work}
\input{figs/data_statistics}

We provide a brief review of relevant fields, including fairness learning in medical imaging, federated learning, and fairness in federated learning.

\subsubsection{Fairness Learning in Medical Imaging}
Fairness learning in medical imaging aims to reduce biases and ensure equitable outcomes for all patient groups, particularly underrepresented minorities. Currently, most research focuses on fairness models and datasets in classification tasks, with limited attention given to cross-domain fairness. Publicly available datasets, such as CheXpert~\cite{irvin2019chexpert}, MIMIC-CXR~\cite{johnson2019mimic}, and Fitzpatrick17k~\cite{groh2021evaluating}, support fairness studies but often lack comprehensive identity attributes, focus predominantly on 2D images, and do not include federated clients/sites representing diverse demographic distributions, limiting their applicability to federated learning tasks or 3D medical imaging. Recent methods~\cite{tian2024fairdomain,tian2024fairseg,jiang2023fair,luo2024fairclip,luo2023harvard,luo2023glau_fair,zong2022medfair,shi2023equitable,shi2024equitable,luo2025fairdiffusion} have made progress in algorithmic fairness for medical imaging, but these approaches mainly address bias within individual clients or sites. The performance of multiple clients in a federated learning setup remains largely unexplored, representing a significant gap in fairness research.

\subsubsection{Federated Learning (FL)} 
FL is a decentralized machine learning paradigm enabling multiple clients to collaboratively train a global model while keeping their local data private. FL approaches typically fall into three categories: traditional fully parameter-updated models, prompt learning methods and LoRA-based methods. 
Traditional FL methods~\cite{mcmahan2017fedavg,li2020federated_survey,fedh2l,fedalign,liu2022few} aggregate model parameters from distributed clients to update a global model, allowing collaborative learning while preserving low communication. FedAvg~\cite{mcmahan2017fedavg} averages local model updates, providing a straightforward and effective solution for various tasks. In contrast, FedProx~\cite{li2020federated_survey} adds a proximal term to the objective function, improving stability and performance in heterogeneous environments. Prompt learning methods ~\cite{zhang2020side, guo2023promptfl,zhao2022reduce,su2022cross,fedotp} customize task-specific text prompts for each client, enabling local and global communication without altering model parameters. PromptFL~\cite{guo2023promptfl} allows clients to train soft prompts instead of the entire model, significantly reducing aggregation overhead and accelerating local training. FedOTP~\cite{fedotp} balances global consensus and local personalization by learning both global and local prompts, using Optimal Transport to align local visual features with prompts and address heterogeneities such as label and feature shifts. The latest methods~\cite{yi2023fedlora,sun2024FFALoRA} introduce LoRA~\cite{hu2022lora} into foundation models to achieve a balance between performance and communication cost.

\subsubsection{Fairness in Federated Learning} 
Fairness in FL \cite{li2019fair} has gained attention due to its unique challenges compared to centralized learning. Research on fairness in FL can be categorized into two main areas: \textit{Site Fairness} and \textit{Group Fairness}, as illustrated in Fig. \ref{fig:site_group}. \textit{Site Fairness} \cite{pan2024fedlf,cheng2024fedgcr,pan2024towards} ensures equitable handling of model updates from clients, especially when data quality, quantity, or distribution varies. Disparities can lead to models favoring clients with larger or higher-quality datasets. Techniques have been developed to equalize client influence during model aggregation, such as FedLF \cite{pan2024fedlf}, which uses multi-objective optimization to minimize gradient conflicts and promote equitable model improvements. \textit{Group Fairness} \cite{ezzeldin2023fairfed,agrawal2024no,badar2024fairtrade} focuses on equitable model performance across demographic groups within clients' local datasets. For example, \cite{ezzeldin2023fairfed} implements a fairness-aware aggregation method, enabling local debiasing and adjusting aggregation weights based on local and global fairness assessments. \cite{badar2024fairtrade} employs multi-objective and Bayesian optimization to balance fairness and accuracy. However, the above existing studies focus on non-medical applications, where biased outcomes are less critical. Despite the urgent need for fairness in healthcare, research on group fairness in federated learning for medical applications is limited, revealing a significant gap in this field.

%% file: figs/data_statistics.tex
\begin{figure*}
    \centering
    \begin{subfigure}{1\textwidth}
        \centering
        \includegraphics[width=0.95\linewidth]{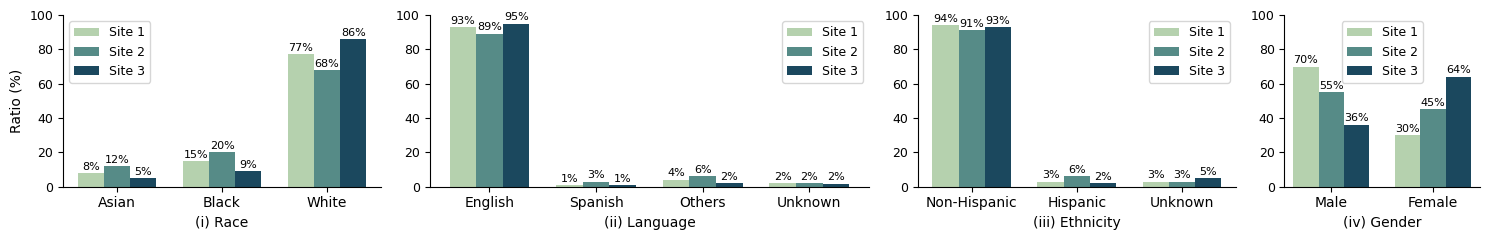}
        \vspace{-2mm}
        \caption{\small The demographic distribution of \revise{ FairFedMed-Oph} across three FL sites}
    \end{subfigure}
    \begin{subfigure}{0.48\textwidth}
        \centering
        \includegraphics[width=0.845\linewidth]{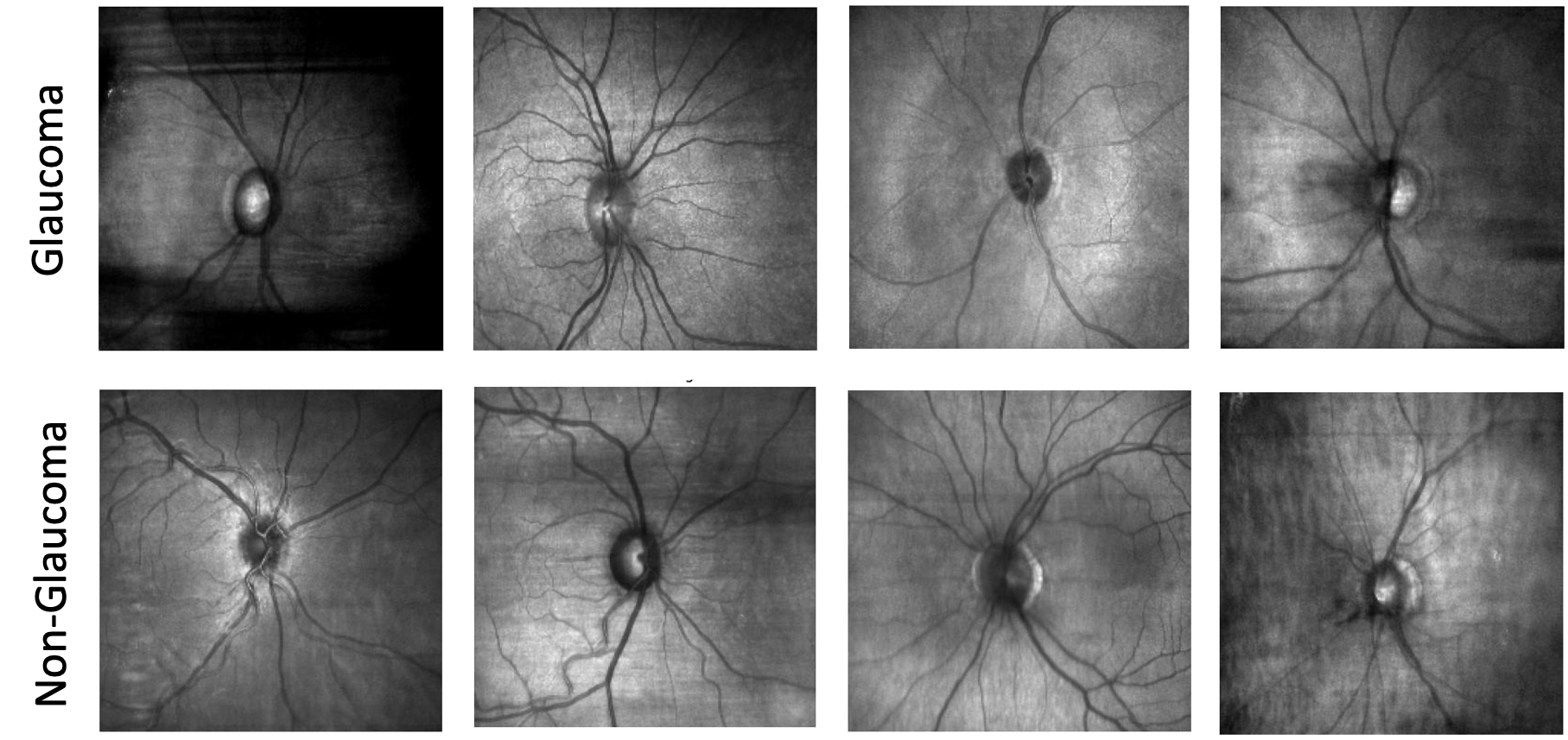}
        \vspace{-1mm}
        \caption{\small 2D SLO fundus images in \revise{ FairFedMed-Oph}}
    \end{subfigure}
    \begin{subfigure}{0.48\textwidth}
        \centering
        \includegraphics[width=0.82\linewidth]{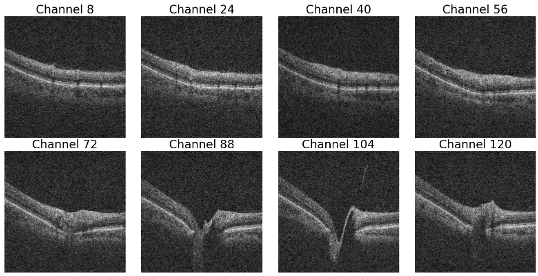}
        \vspace{-1mm}
        \caption{\small 3D OCT B-scan slices in \revise{ FairFedMed-Oph}}
    \end{subfigure}
    \centering
    \begin{subfigure}{0.6\textwidth}
        \centering
        \includegraphics[width=0.95\linewidth]{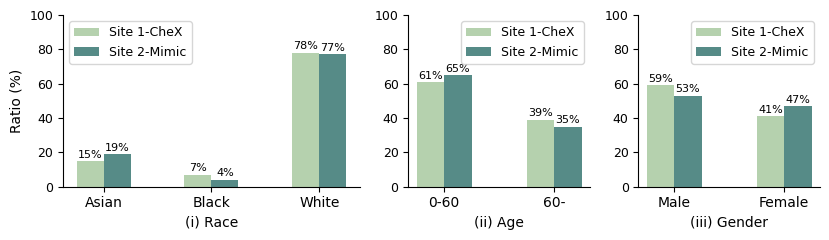}
        \vspace{-2mm}
        \caption{\small \revise{ The demographic distribution of FairFedMed-Chest across two FL sites}}
    \end{subfigure}
    \begin{subfigure}{0.37\textwidth}
        \centering
        \includegraphics[width=0.95\linewidth]{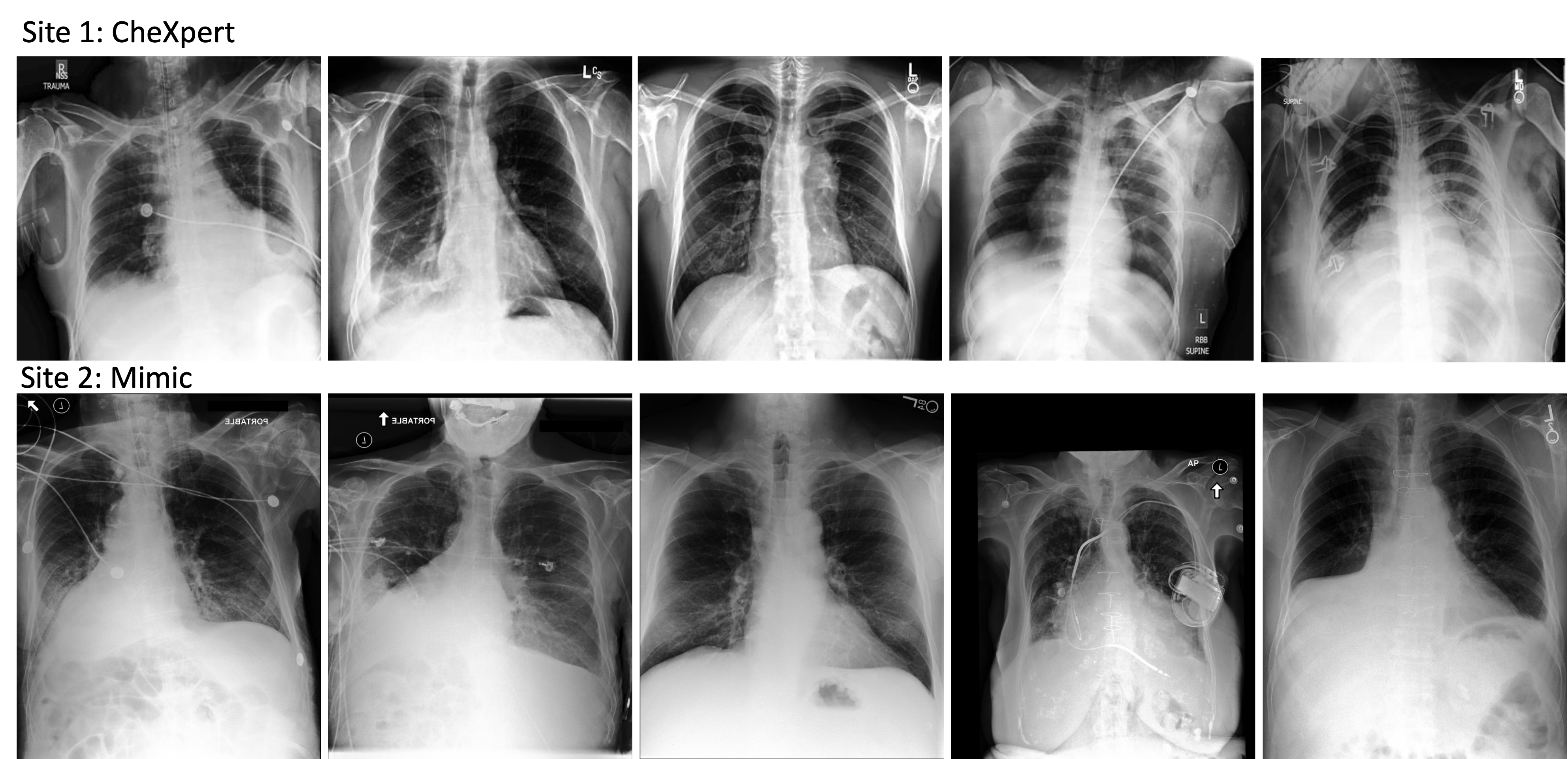}
        \caption{\small \revise{ Chest X-ray images in FairFedMed-Chest}}
    \end{subfigure}
    \vspace{-1mm}
    \caption{\revise{ FairFedMed dataset statistics, covering both FairFedMed-Oph and FairFedMed-Chest. Subfigures (a)–(c) illustrate the demographic distributions and example 2D fundus and 3D OCT images from FairFedMed-Oph, which simulates a FL environment. Subfigures (d)–(e) present FairFedMed-Chest, where subsets of the CheXpert and MIMIC-CXR chest X-ray datasets are treated as two distinct sites to reflect a real-world FL setting.}} \label{fig:three_figs}
\end{figure*}

%% file: secs/3_data_setup.tex
\section{Dataset Analysis}
To facilitate fairness-aware federated learning in the medical domain, we introduce FairFedMed, a comprehensive dataset suite comprising two components: FairFedMed-Oph, which focuses on ophthalmology, and FairFedMed-Chest, which centers on chest X-ray imaging. These datasets are designed to reflect both simulated and real-world federated learning environments, covering diverse medical modalities and demographic attributes. In this section, we provide a detailed analysis of each dataset, including their collection protocols, demographic statistics, and FL configurations.

\subsection{\revise{ FairFedMed-Oph}}
\subsubsection{Data Collection and Quality Control}
To address the shortage of medical datasets for fairness-aware FL, we propose the FairFedMed dataset. The dataset was collected from a large eye hospital, Massachusetts Eye and Ear (MEE) in the United States, which represents populations from a wide range of demographic groups. This study strictly adheres to the principles outlined in the Declaration of Helsinki and has been approved by our institute's Institutional Review Board. All patient subjects in this dataset are de-identified. We selected patients based on the following criteria: 1) All patients received eye care services from MEE between 2010 and 2022. Only one sample per patient was included; 2) All fundus image and OCT scans were reliable with their signal strength not smaller than 6; 3) Images from the last visit of each patient were randomly selected from either the left or right eye. Finally, the dataset contains \revise{ 16,681} samples from \revise{ 16,681} subjects with an average age of 61.3 $\pm$ 16.3 years. Each sample includes both 2D Scanning Laser Ophthalmoscopy (SLO) fundus images and 3D Optical Coherence Tomography (OCT) B-scans, \revise{which are acquired concurrently} using Cirrus devices (Carl Zeiss Meditec, Dublin, California). Each OCT sample contains 128 B-scan images. An example is provided in Fig. \ref{fig:three_figs} (b) and (c). 

\revise{ All images in our dataset have undergone a quality check by board-certified ophthalmologists to ensure they are diagnostically useful. While some images do contain artifacts, we intentionally chose not to exclude all low-quality or artifact-containing images, as our goal is to reflect real-world clinical conditions, where such imperfections are common and unavoidable. We believe that retaining these samples promotes the development of models that are more generalizable and robust in practical deployment scenarios.}

\subsubsection{Data Characteristics}
Within this dataset, we have six demographic attributes including age, gender, race, ethnicity, preferred language, and marital status. The demographic distributions are as follows: \revise{\textit{Age:} $<$60: 40.2\%, and $\geq$60: 59.8\%;}; \textit{Gender:} Female: 57.0\%, and Male: 43.0\%; \textit{Race:} Asian: 8.4\%, Black: 14.7\%, and White: 76.9\%. \textit{Ethnicity:} Non-Hispanic: 92.7\%, Hispanics: 3.7\%, Unknown: 3.5\%. \textit{Preferred Language:} English: 92.5\%, Spanish: 1.6\%, Others: 4.0\%, and Unknown: 1.9\%. \textit{Marital Status:} Married or Partnered: 57.6\%, \revise{Single}: 26.1\%, Divorced: 6.8\%, Legally Separated: 0.9\%, Widowed: 6.1\%, and Unknown: 2.4\%. The glaucoma status of subjects is defined based on a reliable visual field (VF) test, which includes a fixation loss of $\leq$ 33\%, false-positive rate of $\leq$ 20\%, and false-negative rate of $\leq$ 20\%. \revise{A VF test conducted within 30 days of obtaining a fundus image will be utilized to identify glaucoma patients. Criteria for diagnosis include a VF mean deviation less than -3 dB, coupled with abnormal results on both the glaucoma hemifield test and the pattern standard deviation.} The glaucoma and non-glaucoma samples account for 49.0\% and 51.0\%, respectively. 

\subsubsection{Federated Learning Simulation}
\revise{ To simulate the FL environment in this work, we divide all subjects into three separate sites. Specifically, we first analyzed the demographic composition of the full dataset and then introduced slight but meaningful deviations in subgroup proportions across the sites to induce distribution shifts related to demographic attributes. This design better reflects the non-i.i.d. conditions typically encountered in real-world federated learning scenarios. At each site, 70\% of the data is used for training, 10\% for validation, and 20\% for testing.} Note that although three independent sites are studied in this work, it is flexible to simulate more FL sites based on our dataset. \revise{ We adopted a controlled random sampling process to separate the entire dataset into three subsets, ensuring that each subset contains a similar number of samples while introducing distributional differences to simulate realistic FL settings.} We focus on \revise{ four demographic attributes including gender, race, language and ethnicity} with varying cross-site distributions. 
We aim to maximize demographic diversity across different sites, although it is challenging as the majority of participants are White, non-Hispanic, and English-speaking. The demographic distributions across three sites are summarized in Fig. \ref{fig:three_figs} (a). Given the comprehensive image modalities and demographic attributes of subject samples, our dataset can be used to study different FL settings, \textit{i.e.,} varying number of sites and cross-site image modality differences.

\subsection{\revise{ FairFedMed-Chest}}
\revise{
To emulate a real-world FL scenario, we construct the FairFedMed-Chest dataset by treating CheXpert~\cite{irvin2019chexpert} and MIMIC-CXR~\cite{johnson2019mimic} as two distinct clients, representing data from different institutions. This setup captures the challenges of cross-institutional FL, including distributional shifts, demographic discrepancies, and domain-specific variations in chest X-ray imaging protocols. Each site contributes independent patient populations and imaging standards, providing a realistic benchmark for evaluating the robustness and fairness of FL algorithms in medical imaging tasks.
}

\subsubsection{\revise{ Data Characteristics.} }
\revise{
We construct the FairFedMed-Chest dataset by sampling 10,000 chest X-ray images from each of the CheXpert~\cite{irvin2019chexpert} and MIMIC-CXR~\cite{johnson2019mimic} datasets, with an 80\%/10\%/20\% split into training, validation and testing sets for each site. To support our focus on group fairness, the sampling is conducted in a stratified manner to ensure consistent disease prevalence across sites. Specifically, both the CheXpert and MIMIC-CXR subsets are constructed to maintain a uniform disease prevalence of 40.5\%. This design helps eliminate confounding effects from prevalence imbalance, enabling a fairer comparison of model performance across demographic groups and sites. We focus on three key demographic attributes: \textit{Race, Age, and Gender}. As illustrated in Fig.~\ref{fig:three_figs} (d), the two sites show noticeable demographic differences. Site 1 (CheXpert) has a higher proportion of White and younger patients, whereas Site 2 (MIMIC-CXR) includes more Asian individuals and a greater percentage of female patients. These cross-site demographic disparities provide a realistic and challenging setting for evaluating the fairness and robustness of FL algorithms in medical imaging.
}

%% file: secs/4_method.tex
\section{Methodology}

\input{figs/overall}

\subsection{Preliminary}
\revise{ 
In FL, the model is trained collaboratively across multiple clients. Each client maintains a local copy of the model weights, and updates its local model based on its own data. After the local updates, the clients communicate their updated weights to the central server, which aggregates them to update the global model. 
LoRA (Low-Rank Adaptation)~\cite{hu2022lora,meng2024pissa,wang2024milora} is a parameter-efficient fine-tuning method that injects trainable low-rank matrices into pre-trained model weights, allowing adaptation with minimal parameter updates. It significantly reduces memory and computation costs while maintaining performance, making it ideal for large-scale models.
}

\revise{ 
To enhance effective finetuning, we adapt a LoRA within the FL framework, enabling more effective and efficient parameter updates across decentralized datasets. For each client $k$, the model weights incorporating low-rank adaptation at round $t$ can be expressed as: {\footnotesize $\mW^t_k = \mW_0 + \Delta \mW^t_k$}, where {\footnotesize $\mW_0$} is the pre-trained model weights and {\footnotesize $\Delta\mW_k^t$} is the low-rank update in the client $k$ with local data $D_k$. The low-rank term can be implemented either using LoRA~\cite{hu2022lora} (\emph{i.e.}, {\footnotesize $\Delta \mW^t_k=\mU_k^t{\mV_k^t}$}) or SVD-based LoRA~\cite{golub1971svd} (\emph{i.e.}, {\footnotesize $\Delta \mW^t_k=\mU_k^t \mS_k^t {\mV_k^t}$}), where {\footnotesize $\mU_k \in \mathbb{R}^{ m \times r }$} and {\footnotesize $\mV_k \in \mathbb{R}^{ r \times n }$} contain the left and right singular vectors corresponding to the largest $r$ singular values in {\footnotesize $\mS_k \in \mathbb{R}^{r \times r}$}. Both local and global model updates {\footnotesize $\Delta \mW^{t}_k$} and {\footnotesize $\Delta \overline{\mW}^{t}$} are performed using LoRA, formulated as:
}

{\small
\begin{align}
    & \textbf{Local update:} \quad \Delta \mW^{t}_k = \Delta \overline{\mW}^{t-1} - \eta \nabla \mathcal{L}_{D_k}\left(\Delta \overline{\mW}^{t-1}\right),\ \forall k; \nonumber \\
    & \textbf{Global update:} \ \Delta \overline{\mW}^{t} =  \sum\nolimits_{k=1}^K \alpha_k \Delta\mW_k^{t}. \label{eq:lora_global}
\end{align} 
}

\revise{ 
\noindent Here, the learning rate $\eta$ controls the step size of the local optimization. The weight $\alpha_k$ determines the contribution of client $k$ to the global update, typically set based on the relative data size of each client.
This adaptation enables the model to leverage low-rank approximations for the local data at each client, improving computational efficiency while preserving the integrity of the learned representations.
}
\subsection{FairLoRA}\label{sec:method}
\revise{ 
Considering the inherent demographic factors in medical data, the primary challenge is to ensure that the federated model facilitates equitable training across diverse demographic groups. To address this issue, we propose FairLoRA, whose overview is shown in Fig.~\ref{fig:overall}. 
Unlike previous LoRA variants, FairLoRA is a group fairness-aware FL framework for disease classification. It introduces a structured low-rank adaptation mechanism by customizing singular value matrices for each demographic group, effectively preserving intra-group characteristics and reducing bias. At the same time, it shares singular vector matrices across all groups to promote global knowledge transfer and capture inter-group relationships. This design ensures that FairLoRA balances both performance and fairness in medical FL scenarios.
}

The proposed FL framework is depicted on the left side of Fig.~\ref{fig:overall}, where each client trains its own local model, and only the FairLoRA-specific weights are aggregated and updated on the central server. As shown in Fig.~\ref{fig:overall}(a), the model is built upon the CLIP foundation model, with both the text and image encoders kept frozen during training. Its input is an medical disease text prompt (e.g., `Glaucoma'), while the FairLoRA module is embedded in the image encoder to adapt the model for medical images, supporting both 2D fundus and 3D OCT images. The final diagnosis is determined by computing the similarity between the text and image embeddings. The architecture of the FairLoRA module (Fig.~\ref{fig:overall} (b)) can be formulated as:
\begin{align}
    \revise{
    \Delta \mW_k^t = \mU_k^t \mS^t_k {\mV^t_k},\quad \text{\small where} \quad \mS^t_k = \sum\nolimits_{g\in\mathcal{G}}\pi_g \mS^t_{k,g}.}
\end{align}
\revise{Here, $\mS^t_k$ denotes the aggregated singular value matrix at client $k$, and each $\mS^t_{k,g}$ represents the group-specific singular value matrix for demographic group $g$} (e.g., $g \in \mathcal{G}=\{\text{`Black', `Asian', `White'}\}$), preserving the unique characteristics. The vector {\footnotesize $[\pi_1, \cdots, \pi_g, \cdots, \pi_{|\mathcal{G}|}]$} is a one-hot encoding, ensuring that only the target demographic group has a non-zero value, as $\sum_{g\in\mathcal{G}} \pi_g = 1$. The left and right singular vector matrices {\footnotesize $\mU_k^t$} and {\footnotesize $\mV^t_k$} are shared among all demographic groups, allowing the model to capture inter-group relationships. 

\revise{
We use the one-hot encoding vector $\boldsymbol{\pi}$ to control which group-specific singular value matrix $\mS_{k,g}$ gets updated. During local training, if a sample belongs to group $g$, then $\pi_g = 1$ and all other $\pi_{g'} = 0$ for $g' \neq g$. This ensures that only the corresponding $\mS_{k,g}$ receives gradients during backpropagation, while the others remain unchanged. This allows us to update only the singular value matrix $\mS_{k,g}$ corresponding to the target demographic group, without affecting the parameters of other groups, thereby effectively preserving group-specific characteristics and reducing cross-group interference.
}
In FL framework, the local and global updates of FairLoRA module can be mathematically expressed as:

{\small
\begin{align}
    &\! \textbf{Local update:} \quad \forall k,\ \mU^{t}_k \revise{ \mS^t_k} {\mV^{t}_k} \nonumber \\
    & \qquad \qquad \quad = \overline{\mU}^{t\!-\!1}\revise{ \overline{\mS}^{t-1}}{\overline{\mV}^{t\!-\!1}}\!
    - \eta \nabla \mathcal{L}_{D_k}\left(\overline{\mU}^{t\!-\!1}\revise{ \overline{\mS}^{t\!-\!1}}{\overline{\mV}^{t\!-\!1}}\!\right); \label{eq:fairlora_local} \\
    &\! \textbf{Global update:} \quad \overline{\mU}^{t} = \sum\nolimits_{k=1}^K \alpha_{k} \mU_k^{t}, \quad \overline{\mV}^{t} = \sum\nolimits_{k=1}^K \alpha_{k} \mV_k^{t}, \nonumber \\
    & \quad \qquad \quad \ \revise{ \forall g,\ \overline{\mS}_{g}^{t} = \sum\nolimits_{k=1}^K \alpha_{k} {\mS}_{k,g}^{t}, \quad \overline{\mS}^t = \sum\nolimits_{g\in\mathcal{G}} \pi_g\overline{\mS}_{g}^{t}. }\label{eq:fairlora_global}
\end{align} 
}


\noindent Here, {\footnotesize $\overline{\mU}^t$} and {\footnotesize $\overline{\mV}^t$} denote the averaged left and right singular vector matrices, \revise{ and {\footnotesize $\overline{\mS}_g^t$} is the averaged singular value matrix for group $g$ on all clients, preserving intra-group characteristics while leveraging broader samples to reduce bias. The final matrix {\footnotesize $\overline{\mS}^t$} is then computed by aggregating {\footnotesize $\overline{\mS}_g^t$} across all groups using their population proportions $\pi_g$, capturing both intra-group distinctiveness and inter-group fairness. This formulation ensures the global model reflects both client data distributions and demographic group proportions.} The full algorithm is presented in Algorithm \ref{Algorithm}.

\textbf{FairLoRA Initialization.} The initialization of the low-rank matrices in FairLoRA follows a structured approach to align with the SVD framework. The left singular vector matrix {\footnotesize $\overline{\mU}^0=0$} is initialized to zeros to provide a neutral starting point for local updates. The right singular vector matrix {\footnotesize $\overline{\mV}^0 \sim \mathcal{N}(0,1)$} is initialized using a normal distribution, allowing for a diverse range of values to support efficient convergence during training. In addition, the singular value matrices {\footnotesize $\{\overline{\mS}_g^0\}_{g\in\mathcal{G}}$} is initialized using a linear space of values ranging from 0.5 to 0.1, as shown in Fig. \ref{fig:usv}. For all groups, the first half of the ranks is initialized uniformly with this same linear space, ensuring that different demographic groups share the same principal singular vectors. For the remaining half of the ranks, a cyclic pattern is used to initialize the singular values, ensuring that each demographic group has its strongest response at distinct ranks. This design preserves group-specific diversity while enhancing the fairness across different groups. 

\input{secs/algo1}
\input{figs/usv}

\revise{
\textbf{Multi-sensitive Attributes.} Since demographic group information $g \in \mathcal{G}$ is required as part of the model input, our current design supports encoding only one attribute (e.g., race or language) at a time. To extend the model for multiple sensitive attributes, a straightforward solution is to generalize the group-specific singular value matrices $\{\mS_g\}_{g \in \mathcal{G}}$ to $\{\mS_{g^a}\}_{{g^a} \in \mathcal{G}^a}$, where $\mathcal{G}^a$ denotes the demographic groups in the attribute $a$. Since this work primarily focuses on group fairness with respect to a single demographic attribute, we leave the multi-attribute extension for future work, which will further investigate both intra- and inter-attribute relationships.
}

%% file: figs/overall.tex
\begin{figure*}[!t]
    \centering
    \includegraphics[width=0.95\linewidth]{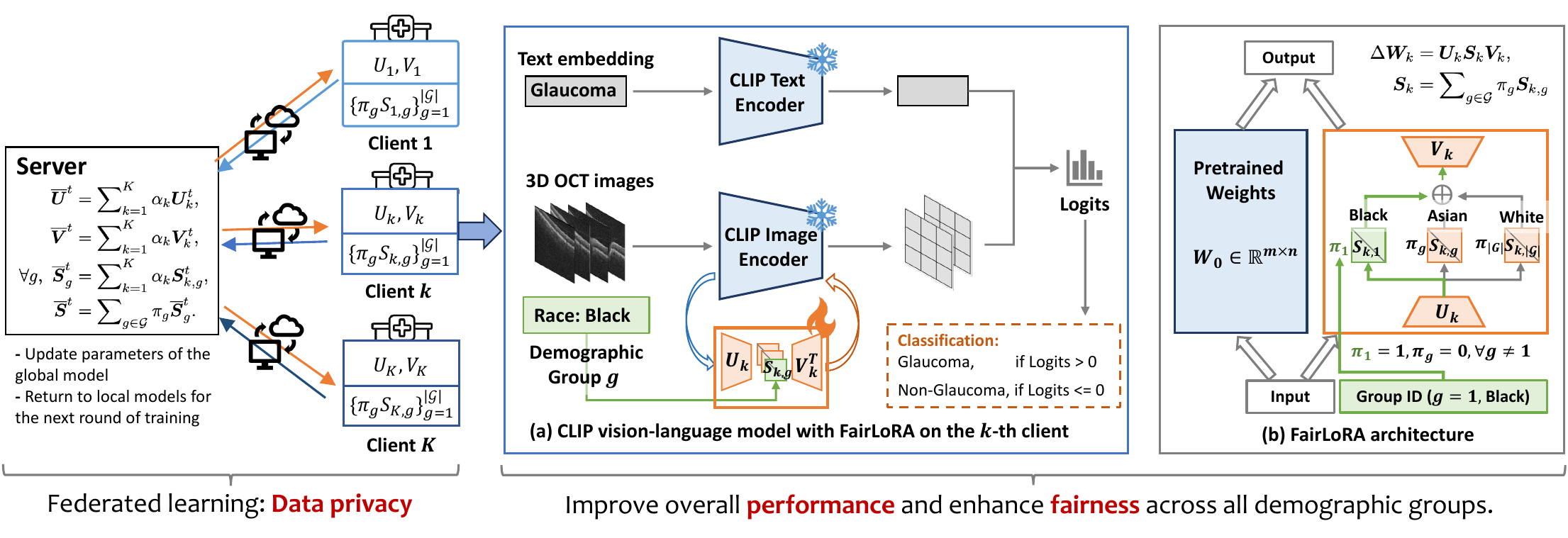}
    \vspace{-2mm}
    \caption{Overview of our proposed group fairness-aware FL framework: \textit{FairLoRA}. Specifically, each client runs a local model, while only the corresponding model weights of \textit{FairLoRA} are globally updated on the server. As shown in the subfigure (a), the model is built on the CLIP foundation model, with the text and image encoders fully frozen. Its input is a medical disease text prompt (e.g., `Glaucoma'), while the \textit{FairLoRA} module is embedded in the image encoder to adapt the model for medical images, supporting both 2D fundus and 3D OCT images. The final diagnosis is determined by computing the similarity between the text and image embeddings. The subfigure (b) shows the core design of \textit{FairLoRA} module: customized singular value matrices $\mS_{k,g}$ for each demographic group (e.g., $\mathcal{G}=\{\text{`Black', `Asian', `White'}\}$), shared singular vector matrices {\footnotesize $\mU_k, \mV_k$} across all demographic groups. Here, $k$ and $g$ denote the client and group indexes, respectively.}
    \label{fig:overall}
\end{figure*}

%% file: secs/algo1.tex
\begin{algorithm}[!t]
    \caption{FairLoRA}\label{Algorithm}
    \begin{algorithmic}[1]
        \STATE \textbf{Input:} 
        \STATE \quad {\footnotesize $K$}: number of clients, $\mathcal{G}$: set of demographic groups
        \STATE \quad $\alpha_k, \alpha_{k,g}$: hyperparameters for each client/group 
        \STATE \quad {\footnotesize $\{D_k\}_{k=1}^K$}: local datasets for each client
        \STATE \quad {\footnotesize $\mW_0$}: pre-trained CLIP model weights
        \STATE \textbf{Initialization:} 
        \STATE \quad Global weights: {\footnotesize $\overline{\mU}^0=\mathbf{0}$, $\overline{\mV}^0 \sim \mathcal{N}(0,1)$, \revise{\{ $\overline{\mS}_{g}^0\}_{g\in \mathcal{G}} \rightarrow \overline{\mS}^0$}}
        
        \FOR{each round $t = 1, 2, \ldots, T$}
            \FOR{probability select clients $k \in \{1, 2, \ldots, K\}$}
                \FOR{each iter $i = 1, 2, \ldots, |D_k|$}
                    \STATE \quad { \revise{Update local weights {\footnotesize $\mU_k^t, \mV_k^t, \mS_k^t$} via Eq. (\ref{eq:fairlora_local})}}
                \ENDFOR
            \ENDFOR
            \STATE { \revise{Update global weights {\footnotesize $\overline{\mU}^t, \overline{\mV}^t, \{\overline{\mS}_g^t\}_{g\in \mathcal{G}}, \overline{\mS}^t$}via Eq. (\ref{eq:fairlora_global})}}
        \ENDFOR
        \STATE \textbf{Output:} \revise{Global weights \ {\footnotesize $\overline{\mU}^T, \overline{\mV}^T, \{\overline{\mS}_g^T\}_{g\in \mathcal{G}}, \overline{\mS}^T$}}
    \end{algorithmic}
\end{algorithm}

%% file: figs/usv.tex
\begin{figure}[t!]
    \centering
    \includegraphics[width=0.99\linewidth]{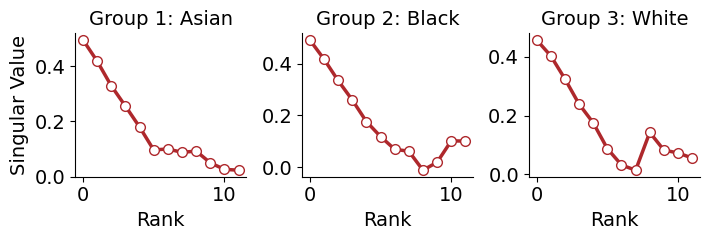}
    \vspace{-6.5mm}
    \caption{Visualization of group fairness-aware singular value matrices $\{\overline{\mS}_g\}_{g\in\mathcal{G}}$ with rank $r = 12$, where the groups of the race attribute is $\mathcal{G}=\{\text{`Asian', `Black', `White'}\}$.}\label{fig:usv}
\end{figure}

%% file: secs/5_experiments.tex
\section{Experiments}

\input{tables/table1_slo_vitb}
\input{tables/table2_slo_r50}

\subsection{Experimental Setup}

\subsubsection{Implementation Details} To ensure a focused analysis while avoiding unnecessary complexity, we select race, ethnicity, preferred language \revise{ and gender} as the key demographic attributes in this study. These attributes are widely recognized in fairness-related research and capture essential demographic variations that may impact model performance. By focusing on these three factors, we ensure a more interpretable and targeted evaluation of fairness-aware federated learning.

The training process consists of 50 epochs with a batch size of 32. The optimization is performed using Stochastic Gradient Descent (SGD) with an initial learning rate of 0.001, which is reduced by a factor of 0.1 at epoch 40. In each training round, two out of three sites are randomly selected to update their local model weights, which are then aggregated to update the global model. To ensure training stability, we employ the exponential moving average (EMA)~\cite{morales2024exponential} strategy to update the global parameters. \textit{FairLoRA} is implemented with two representative backbone architectures: ResNet50 \cite{he2016deep} and ViT-B \cite{dosovitskiy2020image}. The LoRA rank and alpha parameters are set to (12, 2) for the ViT-B backbone and (32, 8) for the ResNet50 backbone. Additionally, all batch normalization layers in the ResNet50 architecture are set to trainable.

\revise{ In the 3D setting, we uniformly sample 8 representative slices from each volumetric OCT scan, which originally contains 128 slices. To process each 3D slice, we apply a dedicated projection module that maps the multi-channel slice representation into a standard 3-channel format using a convolutional layer with a $5 \times 5$ kernel and appropriate padding to preserve spatial dimensions. }

\subsubsection{Baselines} We compare \textit{FairLoRA} with several representative FL baselines, including fully parameter-updated, prompt-based and adapter-based models. \textit{FedAvg}~\cite{mcmahan2017fedavg} performs full parameter averaging across clients but ignores fairness. \textit{FedHEAL}~\cite{chen2024fedheal} improves fairness under domain skew via selective local updates based on parameter relevance. \textit{PromptFL}~\cite{guo2023promptfl} trains soft prompts instead of full models for efficient and privacy-preserving FL, while \textit{FedOTP}~\cite{fedotp} adopts Optimal Transport to coordinate global-local prompt learning across clients—however, both are fairness-agnostic. \textit{ViTAdapter}~\cite{chen2022vitadapter} applies lightweight adapters to vision transformers for efficient FL, yet does not consider fairness. In contrast, \textit{FairLoRA} incorporates group-aware fairness by customizing singular value matrices for each demographic group while sharing singular vector matrices globally, achieving both equitable representation and model efficiency.

\subsubsection{Evaluation Metrics}
We evaluated the performance of all baseline models and our proposed model using the Overall Area Under the Curve (AUC), Equality-Scale AUC (ES-AUC)\cite{luotmi}, and Group-wise AUCs, along with fairness metrics including Equal Opportunity Difference (EOD)\cite{hardt2016equality} and Statistical Parity Difference (SPD)~\cite{dwork2012fairness}.
\revise{ Note that while EOD and SPD are widely used fairness metrics, they were originally designed for binary group comparisons and may be less informative in settings involving more than two demographic groups, such as race. In contrast, ES-AUC captures performance consistency across all subgroups, offering a more reliable and interpretable measure of group fairness in multi-group scenarios.
}

\input{tables/table3_oct_vitb}
\input{tables/table4_oct_r50}

\subsection{Results on \revise{ FairFedMed-Oph}}

\subsubsection{Results on 2D SLO Fundus Images}
Tables~\ref{tab:vit_slo} and \ref{tab:r50_slo} present the performance and fairness of various FL models for glaucoma detection on FairFedMed-Oph using 2D SLO fundus images. 
While fully parameter-updated models \textit{FedAvg} and \textit{FedHEAL} achieve high overall AUCs (73–78\%) on 2D SLO images, their fairness metrics (ES-AUC 65–69\%) vary widely, indicating poor demographic fairness despite good accuracy. \revise{ Prompt-based and adapter-based and  models (\textit{PromptFL}, \textit{FedOTP} and \textit{ViTAdapter}) generally achieve lower overall AUCs (71–76\%), primarily because they rely on frozen CLIP backbones pretrained on natural images, which are not well-suited for medical imaging tasks. However, these models improve group fairness, reducing EOD and SPD by over 10\%, showing the advantage of generalizable representations despite lower domain-specific performance.}


Our \textit{FairLoRA} outperforms other FL models by achieving the highest overall AUC and ES-AUC, while also performing well on fairness metrics such as EOD and SDP. Though EOD and SDP are more sensitive, ES-AUC offers a more holistic measure of group fairness. For example, on the race attribute with a ViT-B backbone, \textit{FairLoRA} improves overall AUC and ES-AUC by 3.1\% and 2.5\% over the next-best method. While its EOD and SDP are slightly higher than those of prompt-based models, they remain competitive, reflecting a balance between performance and fairness. Similar patterns hold across other attributes and backbones.

\subsubsection{Results on 3D OCT B-Scan Images}
\revise{Tables~\ref{tab:vit_oct} and \ref{tab:r50_oct}} show that while traditional FL methods (\textit{FedAvg}, \textit{FedHEAL}) achieve strong AUCs (74\%–78\%), their fairness metrics vary widely. \revise{ Prompt-based and adapter-based models (\textit{ViTAdapter}, \textit{PromptFL} and \textit{FedOTP}) slightly trail in AUC but offer improved fairness.} \textit{FairLoRA} consistently outperforms all baselines, with 3–7\% gains in overall AUC and significantly higher ES-AUC, while maintaining competitive EOD and SDP.
\revise{ 
In summary, \textit{FairLoRA} achieves a strong balance between performance and fairness, outperforming existing FL models. While traditional FL methods offer high accuracy, they lack fairness across demographic groups. Prompt-based and adapter-based models improve fairness but lose performance due to limited adaptability to medical data. \textit{FairLoRA} maintains high classification accuracy while significantly enhancing fairness across all demographic groups, making it a promising solution for fairness-aware medical FL.}

\subsection{\revise{ Results on FairFedMed-Chest}}
\revise{
In the real-world FL setting using CheXpert and MIMIC as two sites, \textit{FairLoRA} significantly outperforms prompt-based and adapter-based baselines in both performance and fairness evaluation. As shown in Tables~\ref{tab:vit_chest} and ~\ref{tab:r50_chest}, \textit{FairLoRA} achieves the highest overall AUC (82.6\% and 84.1\%) and ES-AUC (78.6\% and 81.4\%) on the race attribute, outperforming \textit{PromptFL} and \textit{FedOTP} by 6-10\%. Similar trends are observed for gender and age attributes, with \textit{FairLoRA} improving both average and subgroup AUCs, while reducing fairness gaps (EOD and SPD) by a large margin, achieving up to 3-6\% reductions compared to baselines. 
Compared to the simulated three-site setting in FairFedMed-Oph, the real-world two-institution setup in FairFedMed-Chest better highlights the effectiveness of \textit{FairLoRA} in promoting group fairness. This demonstrates that \textit{FairLoRA} is particularly well-suited for realistic FL scenarios, where it consistently balances strong performance with enhanced fairness across demographic groups.
}

\input{tables/table5_vit_chexmimic}
\input{tables/table6_r50_chexmimic}

\subsection{Ablation Study}
\input{figs/abl_study}
\subsubsection{Local vs. Federated Learning} 
Fig.~\ref{fig:single_federated} compares the performance of \textit{FairLoRA} under local learning and federated learning across three clients. Federated learning consistently outperforms local training in terms of overall AUC, ES-AUC, and group-wise AUCs, with average gains of 1–3\%. These improvements demonstrate that integrating shared knowledge from other clients enables \textit{FairLoRA} to generalize better across diverse data distributions and demographic subgroups.
\revise{ 
Federated learning enhances \textit{FairLoRA}'s accuracy and fairness, making it robust and equitable for multi-site medical imaging tasks.
}

\subsubsection{FairLoRA vs. Other LoRA Variants}
Fig.~\ref{fig:LoRAS} presents a comparative evaluation of \textit{FairLoRA} against other LoRA-based variants. Specifically, we compare the baseline model, \textit{PromptFL}, with three approaches that incorporate low-rank adaptation: standard LoRA, SVD-based LoRA, and our proposed \textit{FairLoRA}. \textit{PromptFL} exhibits the lowest overall performance, indicating significant fairness disparities across demographic groups. Incorporating LoRA for fine-tuning improves overall AUC by 3\% and ES-AUC by 6\%, highlighting its effectiveness in boosting model adaptability. The SVD-based LoRA model maintains strong overall AUC performance; however, its ES-AUC and group-wise AUCs fluctuate across clients, indicating that it struggles to maintain fairness under distribution shifts. 
In contrast, \textit{FairLoRA} achieves the best overall performance, with the highest AUC of 79.3\%, along with superior ES-AUC and group-wise AUCs, demonstrating its effectiveness in improving classification accuracy while ensuring group fairness.
\revise{ 
\textit{FairLoRA} proves to be the most effective LoRA variant, achieving both high performance and equitable outcomes across diverse demographic groups.
}

\subsubsection{Training Convergence of LoRAs} Fig.~\ref{fig:lora_convergence} illustrates the overall AUC convergence of different models during training, with all model weights updated using exponential moving average (EMA). 
The LoRA model exhibits significant instability during training, as it fails to effectively capture data distribution variations in the federated setting, leading to large gradient fluctuations. Its AUC fluctuates sharply, making it difficult to maintain consistent performance. SVD-based LoRA achieves a similar AUC but is slightly less stable, as its AUC declines slightly after reaching its peak and eventually stabilizes. 
In contrast, \textit{FairLoRA} demonstrates the most stable and superior performance, maintaining a highest overall AUC after early convergence with minimal fluctuations. 
\revise{ 
These results highlight \textit{FairLoRA}’s robustness and its advantage in ensuring both stable training and high accuracy in medical imaging classification tasks.
}

\subsubsection{Singular Value Initialization}  
Fig.~\ref{fig:s_init_abl} shows the impact of initializing $\{ \overline{\mS}_{g}^0 \}_{g\in \mathcal{G}}$ on overall AUC. Initializing all groups with the same linear values (`S-init 1' in Fig.~\ref{fig:s_init_abl}) yields the highest early AUC (79.5\%) but drops to 77.5\%, likely due to its inability to distinguish group differences. 
In contrast, the cyclic shift initialization method (i.e., dividing singular values into group-sized parts and cyclically assigning the largest to each rank, denoted as `S-init 2' in Fig.~\ref{fig:s_init_abl}) results in a slight increase in overall AUC but with lower convergence efficiency, which may lead to less optimal training.  
The half-half initialization method (i.e., keeping the first half of singular values identical across groups and cyclically shifting the second half, denoted as `S-init 3' in Fig.~\ref{fig:s_init_abl}) performs the most stably, with the AUC consistently around 79\%. 
\revise{ 
FairLoRA balances group influence by ensuring stable global features with the first half of singular values while the cyclic shift in the second half enhances adaptability, improving fairness and robustness for medical imaging tasks.
}

\input{tables/table8_2_out_3}
\subsubsection{Client Selection on Local Training.}
\revise{
To investigate the influence of client participation on performance and fairness, we vary the number of selected clients during local training under the ViT-B backbone. Table~\ref{tab:num_clients_local_training} shows the results of varying the number of participating clients during local training on FairFedMed-Oph. Using 2 out of 3 clients yields slightly lower overall AUC (82.4 vs. 82.7) and results in degraded fairness, as indicated by a lower ES-AUC (77.8 vs. 79.6). These findings indicate that while subsampling clients (e.g., 2 out of 3) slightly impacts subgroup performance and fairness, it maintains comparable overall accuracy. The minimal performance gap suggests that partial participation introduces limited bias, making it a practical solution for reducing computational cost while enabling reliable fairness evaluation. However, for optimal performance and fairness across all subgroups, including all clients in training remains the preferred approach.
}

\input{tables/table7_abl}
\subsubsection{Robustness to Missing Demographic Metadata at Inference} 
\revise{
In scenarios where demographic attributes are unavailable or incomplete during inference, we propose using the overall population distribution to replace the one-hot demographic encoding. That is, each $\pi_g$ can be set to the proportion of group $g$ in the overall dataset. 
Table~\ref{tab:wo_demo_inf} shows that \textit{FairLoRA} remains effective even without demographic metadata during inference. Replacing one-hot group encoding with population proportions causes only a minor drop in performance (e.g., –0.5\% Overall AUC, –1.7\% ES-AUC), while SPD improves by 2.6\%. This demonstrates the model's robustness and practicality in real-world scenarios where demographic attributes may be unavailable.
}

%% file: tables/table1_slo_vitb.tex
\begin{table*}[!t]
    \centering
    \caption{Quantitative performance and fairness evaluation on \textbf{FairFedMed-Oph using 2D SLO Fundus} images, trained with the \textbf{ViT-B/16} backbone. `Avg.' represents the performance of the global model. `Hisp.' is an abbreviation for `Hispanic'. $\uparrow$ indicates higher is better, while $\downarrow$ indicates lower is better. \textbf{Bold} indicates the best results, {underline} denotes the second-highest.}
    \resizebox{0.99\textwidth}{!}{
    \setlength{\tabcolsep}{1.5pt}
    \begin{NiceTabular}{lc|ccccccc|ccccccc|cccccc|cccccc}
    \toprule
       \multicolumn{2}{c|}{\textbf{Attribute}} &\multicolumn{7}{c|}{\textbf{Race}} &\multicolumn{7}{c|}{\textbf{Language}} &\multicolumn{6}{c}{\textbf{Ethnicity}} &\multicolumn{6}{c}{\revise{ \textbf{Gender}}} \\
       \midrule
       Model &\shortstack{Client \\ ID}  &\shortstack{Overall \\ AUC$\uparrow$} &\shortstack{ES \\ AUC$\uparrow$} &\shortstack{Asian \\ AUC$\uparrow$} &\shortstack{Black \\ AUC$\uparrow$} & \shortstack{White \\ AUC$\uparrow$} &EOD$\downarrow$ &SPD$\downarrow$ &\shortstack{Overall \\ AUC$\uparrow$} &\shortstack{ES \\ AUC$\uparrow$} &\shortstack{\!English\! \\ AUC$\uparrow$} &\shortstack{Spanish \\ AUC$\uparrow$} &\shortstack{Others \\ AUC$\uparrow$} &EOD$\downarrow$ &SPD$\downarrow$ &\shortstack{Overall \\ AUC$\uparrow$} &\shortstack{ES \\ AUC$\uparrow$} &\shortstack{NonHisp. \\ AUC$\uparrow$} &\shortstack{Hisp. \\ AUC$\uparrow$} &EOD$\downarrow$ &SPD$\downarrow$ &\shortstack{\revise{ Overall} \\ \revise{AUC$\uparrow$}} &\shortstack{\revise{ ES }\\ \revise{ AUC$\uparrow$}} &\shortstack{\revise{ Male} \\ \revise{ AUC$\uparrow$}} &\shortstack{\revise{ Female} \\ \revise{ AUC$\uparrow$}} &\revise{ EOD$\downarrow$} &\revise{ SPD$\downarrow$} \\
       \midrule
        \multirow{4}{*}{\rotatebox{0}{FedAvg}}
        &C1   &75.7$\pm$0.7&69.1$\pm$1.2&79.3$\pm$4.5&69.9$\pm$1.4&75.5$\pm$0.9&34.2$\pm$2.6&31.8$\pm$2.9&76.2$\pm$0.2&56.6$\pm$0.3&75.2$\pm$0.3&52.8$\pm$2.1&86.5$\pm$3.2&40.7$\pm$0.8&37.2$\pm$1.4&77.1$\pm$0.9&72.3$\pm$1.6&79.2$\pm$1.0&72.6$\pm$0.5&23.8$\pm$1.5&17.5$\pm$1.0&74.2$\pm$1.2& 71.4$\pm$0.0& 75.7$\pm$2.0& 71.8$\pm$0.3& 4.8$\pm$0.2& 3.6$\pm$1.6\\
        &C2   &74.6$\pm$0.3&65.7$\pm$0.8&81.5$\pm$1.0&78.6$\pm$0.4&71.9$\pm$0.6&23.5$\pm$2.0&21.8$\pm$1.4&75.8$\pm$0.6&67.6$\pm$0.4&76.2$\pm$0.6&75.7$\pm$1.3&64.3$\pm$2.3&38.9$\pm$3.0&31.5$\pm$2.1&78.9$\pm$0.1&67.6$\pm$0.1&78.2$\pm$0.1&62.7$\pm$0.0&27.6$\pm$2.5&20.4$\pm$3.3&74.2$\pm$0.4& 72.0$\pm$1.0& 75.8$\pm$0.0& 72.8$\pm$0.8& 3.1$\pm$1.1& 0.7$\pm$0.4\\ 
        &C3   &74.2$\pm$0.3&68.5$\pm$3.0&78.7$\pm$4.4&73.3$\pm$1.1&71.2$\pm$0.8&19.4$\pm$2.4&14.6$\pm$3.5&73.2$\pm$0.1&62.9$\pm$1.8&74.3$\pm$0.1&70.9$\pm$8.0&60.3$\pm$2.2&33.7$\pm$4.5&29.2$\pm$0.1&71.7$\pm$0.1&64.5$\pm$0.7&73.1$\pm$0.0&62.1$\pm$1.1&30.6$\pm$1.1&23.5$\pm$0.7&74.9$\pm$0.6& 71.1$\pm$0.2& 78.6$\pm$1.0& 73.2$\pm$0.4& 4.7$\pm$0.8& 1.3$\pm$0.1\\ 
        \rowcolor{blue!5} \cellcolor{nocolor} &Avg. &74.8$\pm$0.3&67.7$\pm$0.5&79.8$\pm$1.3&73.0$\pm$0.2&{72.9}$\pm$0.3&25.7$\pm$1.8&22.7$\pm$2.5&75.1$\pm$0.1&{62.4}$\pm$1.4&75.3$\pm$0.1&66.5$\pm$2.1&70.4$\pm$0.4&37.7$\pm$1.8&32.6$\pm$0.4&{75.9}$\pm$0.3&68.1$\pm$1.3&76.9$\pm$0.4&65.8$\pm$1.4&27.3$\pm$0.1&20.4$\pm$0.4&74.4$\pm$0.0& 71.8$\pm$0.7& 76.4$\pm$0.6& 72.8$\pm$0.4& 3.0$\pm$2.1& 1.0$\pm$0.1\\ 
        \midrule
        \multirow{4}{*}{\rotatebox{0}{FedHEAL}}
        &C1   &76.7$\pm$0.2&69.2$\pm$0.4&80.2$\pm$0.5&70.3$\pm$0.2&77.6$\pm$0.6&28.2$\pm$2.0&25.7$\pm$1.7&75.3$\pm$0.2&55.8$\pm$0.1&76.3$\pm$0.2&53.5$\pm$0.0&87.6$\pm$0.0&41.4$\pm$3.5&35.6$\pm$2.4&76.9$\pm$0.1&72.5$\pm$0.6&78.2$\pm$0.1&72.2$\pm$1.0&24.7$\pm$4.7&21.5$\pm$1.1&76.3$\pm$0.3& 71.8$\pm$0.3& 78.6$\pm$0.2& 72.3$\pm$0.3& 3.6$\pm$1.1& 1.9$\pm$0.3\\ 
        &C2   &75.1$\pm$0.2&66.3$\pm$0.1&80.3$\pm$0.3&79.6$\pm$0.2&71.5$\pm$0.1&21.4$\pm$1.9&19.6$\pm$0.7&76.5$\pm$0.1&67.3$\pm$0.0&75.5$\pm$0.1&74.9$\pm$0.6&65.3$\pm$0.3&37.9$\pm$0.2&33.9$\pm$0.5&77.5$\pm$0.1&67.5$\pm$0.4&79.7$\pm$0.0&64.9$\pm$0.6&28.3$\pm$1.3&23.8$\pm$0.0&76.1$\pm$0.0& 75.8$\pm$0.1& 76.1$\pm$0.2& 75.9$\pm$0.1& 3.6$\pm$0.1& 1.9$\pm$0.5\\ 
        &C3   &75.9$\pm$0.0&67.8$\pm$0.4&79.8$\pm$1.0&76.3$\pm$0.5&68.2$\pm$0.0&22.8$\pm$2.5&17.3$\pm$1.8&73.9$\pm$0.0&64.1$\pm$0.1&74.3$\pm$0.0&71.5$\pm$1.1&61.4$\pm$1.4&32.5$\pm$1.5&27.8$\pm$1.1&73.1$\pm$0.1&64.9$\pm$0.2&75.2$\pm$0.1&62.5$\pm$0.0&26.2$\pm$5.1&19.6$\pm$4.9&76.1$\pm$0.1& 74.9$\pm$0.1& 77.3$\pm$0.0& 75.7$\pm$0.1& 2.1$\pm$2.0& 1.2$\pm$0.1\\ 
        \rowcolor{blue!5} \cellcolor{nocolor} &Avg. &{75.9}$\pm$0.0 &{67.8}$\pm$0.3&{80.1}$\pm$0.3&\textbf{75.4}$\pm$0.3&72.4$\pm$0.1&24.1$\pm$1.5&20.8$\pm$1.3&{75.2}$\pm$0.1&{62.4}$\pm$0.3&{75.4}$\pm$0.1&66.6$\pm$0.1&{71.4}$\pm$0.6&{37.2}$\pm$2.8&32.4$\pm$0.8&75.8$\pm$0.1&{68.3}$\pm$0.1&{77.7}$\pm$0.1&{66.6}$\pm$0.1&26.4$\pm$2.1&21.6$\pm$1.7&76.2$\pm$0.1& 74.4$\pm$0.2& \textbf{77.5}$\pm$0.0& 75.1$\pm$0.1& \textbf{0.4}$\pm$0.0& \textbf{0.9}$\pm$0.1 \\ 
        
        \midrule
        \multirow{4}{*}{\rotatebox{0}{PromptFL}}
        &C1   &71.9$\pm$0.2& 67.5$\pm$0.4& 77.2$\pm$0.8& 72.6$\pm$0.6& 71.3$\pm$0.3& 13.5$\pm$5.5& 8.7$\pm$1.0&69.9$\pm$0.4& 59.5$\pm$3.6& 69.5$\pm$0.5& 66.7$\pm$5.4& 83.9$\pm$1.5& 27.8$\pm$6.1& 8.4$\pm$0.5&72.9$\pm$0.3& 70.9$\pm$1.2& 72.9$\pm$0.3& 70.1$\pm$1.6& 20.1$\pm$6.1& 16.8$\pm$3.9&72.3$\pm$0.2& 71.8$\pm$0.5& 72.1$\pm$0.3& 72.9$\pm$0.1& 3.0$\pm$0.2& 1.5$\pm$1.5\\
        &C2   &72.6$\pm$0.0& 68.0$\pm$0.1& 76.3$\pm$0.1& 70.1$\pm$0.0& 73.1$\pm$0.1& 15.0$\pm$0.7& 4.2$\pm$0.8&75.0$\pm$0.5& 68.6$\pm$0.6& 74.5$\pm$0.5& 77.2$\pm$0.4& 81.6$\pm$1.1& 27.2$\pm$1.2& 21.1$\pm$0.7&73.6$\pm$0.6& 68.6$\pm$0.9& 73.7$\pm$0.6& 66.5$\pm$1.2& 22.1$\pm$1.6& 16.4$\pm$0.3&74.1$\pm$0.4& 71.0$\pm$1.1& 72.1$\pm$0.9& 76.5$\pm$0.2& 4.9$\pm$0.2& 4.1$\pm$0.5\\
        &C3   &74.7$\pm$0.3& 71.3$\pm$1.6& 72.5$\pm$1.3& 73.2$\pm$2.1& 75.0$\pm$0.1& 9.7$\pm$4.0& 16.4$\pm$2.9&73.2$\pm$0.2& 63.9$\pm$0.2& 72.9$\pm$0.2& 83.3$\pm$0.0& 70.9$\pm$3.3& 33.3$\pm$0.0& 12.4$\pm$2.7&72.6$\pm$0.5& 64.9$\pm$0.3& 72.2$\pm$0.6& 84.0$\pm$1.1& 19.2$\pm$5.3& 8.9$\pm$3.4&73.3$\pm$0.3& 69.3$\pm$0.0& 69.8$\pm$0.0& 75.6$\pm$0.5& 8.4$\pm$0.8& 2.3$\pm$0.1\\
        \rowcolor{blue!5} \cellcolor{nocolor} &Avg. &73.0$\pm$0.2& 68.9$\pm$0.7& 75.3$\pm$0.7& 72.0$\pm$0.9& 73.1$\pm$0.2& 12.7$\pm$3.4& 9.8$\pm$1.6&72.7$\pm$0.4& 64.0$\pm$1.5& 72.3$\pm$0.4& 75.7$\pm$1.9& 78.8$\pm$2.0& 29.4$\pm$2.4& \textbf{14.0}$\pm$1.3&73.0$\pm$0.5& 68.1$\pm$0.8& 73.0$\pm$0.5& 73.5$\pm$1.3& 20.5$\pm$4.3& 14.0$\pm$2.6&73.2$\pm$0.3& 70.7$\pm$0.5& 71.4$\pm$0.4& 75.0$\pm$0.3& 5.4$\pm$0.4& 2.6$\pm$0.7 \\
        \midrule
        \multirow{4}{*}{\rotatebox{0}{FedOTP}}
        &C1  &71.9$\pm$0.3& 69.5$\pm$0.0& 74.8$\pm$0.9& 71.6$\pm$0.5& 71.7$\pm$0.1& 7.2$\pm$0.6& 7.3$\pm$0.4&69.5$\pm$0.3& 58.1$\pm$0.9& 69.1$\pm$0.4& 77.3$\pm$3.2& 81.1$\pm$0.1& 20.8$\pm$4.2& 6.6$\pm$2.2&72.8$\pm$0.0& 67.1$\pm$1.3& 73.0$\pm$0.0& 64.6$\pm$2.1& 22.2$\pm$3.3& 16.0$\pm$3.1&71.6$\pm$0.4& 71.3$\pm$0.5& 71.7$\pm$0.4& 71.4$\pm$0.5& 3.1$\pm$0.5& 2.4$\pm$0.7\\
        &C2   &72.7$\pm$0.6& 68.9$\pm$0.5& 76.4$\pm$0.3& 71.0$\pm$0.2& 72.8$\pm$0.7& 15.6$\pm$1.7& 8.4$\pm$1.6&74.8$\pm$0.3& 68.2$\pm$0.4& 74.2$\pm$0.3& 79.5$\pm$0.1& 79.2$\pm$0.4& 28.9$\pm$1.4& 18.8$\pm$0.5&72.5$\pm$0.1& 69.2$\pm$1.0& 72.5$\pm$0.1& 67.8$\pm$1.5& 24.2$\pm$0.2& 21.0$\pm$0.3&73.3$\pm$0.0& 71.5$\pm$0.1& 72.1$\pm$0.0& 74.5$\pm$0.1& 5.8$\pm$1.7& 4.5$\pm$1.3\\
        &C3   &75.3$\pm$0.1& 70.8$\pm$0.0& 70.5$\pm$0.1& 74.3$\pm$0.2& 75.8$\pm$0.1& 10.7$\pm$0.9& 15.1$\pm$4.8&73.3$\pm$0.6& 64.6$\pm$0.5& 73.0$\pm$0.6& 83.3$\pm$0.0& 76.5$\pm$1.2& 30.3$\pm$7.5& 18.6$\pm$0.1&72.7$\pm$0.1& 63.3$\pm$0.3& 72.4$\pm$0.0& 87.3$\pm$0.4& 13.5$\pm$0.6& 12.0$\pm$0.9&72.9$\pm$0.2& 68.1$\pm$0.2& 68.7$\pm$0.2& 75.6$\pm$0.4& 8.4$\pm$0.2& 3.5$\pm$0.7\\
        \rowcolor{blue!5} \cellcolor{nocolor} &Avg. &73.3$\pm$0.3& 69.7$\pm$0.2& 73.9$\pm$0.4& 72.3$\pm$0.3& 73.4$\pm$0.3& \textbf{11.2}$\pm$1.0& 10.3$\pm$2.2&72.5$\pm$0.4& 63.6$\pm$0.6& 72.1$\pm$0.4& \textbf{80.0}$\pm$1.1& 78.9$\pm$0.6& \textbf{26.7}$\pm$4.4& 14.6$\pm$0.9&72.7$\pm$0.1& 66.5$\pm$0.9& 72.6$\pm$0.1& 73.2$\pm$1.4& 20.0$\pm$1.4& 16.3$\pm$1.4&72.6$\pm$0.2& 70.3$\pm$0.2& 70.8$\pm$0.2& 73.8$\pm$0.3& 5.8$\pm$0.8& 3.5$\pm$0.9\\

        \midrule
        \multirow{4}{*}{\rotatebox{0}{\revise{ViTAdapter}}}
        &\revise{C1}   &74.4$\pm$0.1& 69.6$\pm$0.2& 71.6$\pm$0.4& 72.7$\pm$0.2& 75.4$\pm$0.1& 11.5$\pm$0.1& 10.8$\pm$4.4
        &74.0$\pm$0.2& 66.5$\pm$2.0& 74.9$\pm$0.0& 72.9$\pm$2.8& 65.4$\pm$4.2& 59.1$\pm$17.9& 37.9$\pm$2.9
        &73.8$\pm$0.5& 70.7$\pm$1.3& 74.8$\pm$0.1& 69.5$\pm$1.8& 9.5$\pm$3.1& 1.3$\pm$1.3
        &73.8$\pm$0.2& 72.3$\pm$0.5& 73.2$\pm$0.3& 76.2$\pm$0.3& 5.2$\pm$3.9& 2.1$\pm$1.2\\
        
        &\revise{C2}   &75.5$\pm$0.3& 67.7$\pm$1.0& 81.3$\pm$1.2& 70.8$\pm$1.1& 77.1$\pm$0.2& 17.7$\pm$2.2& 2.7$\pm$1.5
        &75.2$\pm$0.1& 68.1$\pm$2.7& 75.4$\pm$0.4& 66.1$\pm$4.7& 77.1$\pm$0.9& 30.7$\pm$7.9& 15.4$\pm$1.9
        &75.4$\pm$0.1& 73.4$\pm$1.8& 75.3$\pm$0.3& 72.9$\pm$2.6& 19.3$\pm$5.0& 11.2$\pm$2.1
        &75.3$\pm$0.1& 73.0$\pm$0.6& 74.1$\pm$0.4& 77.9$\pm$0.8& 5.8$\pm$2.6& 4.8$\pm$1.9\\
        
        &\revise{C3}   &74.5$\pm$0.6& 66.7$\pm$2.8& 76.8$\pm$5.9& 68.2$\pm$2.8& 75.3$\pm$0.4& 17.3$\pm$0.2& 14.0$\pm$1.0
        &74.1$\pm$0.4& 72.1$\pm$0.9& 74.4$\pm$0.4& 75.8$\pm$3.5& 74.5$\pm$1.1& 45.8$\pm$7.2& 36.2$\pm$14.5
        &74.3$\pm$0.4& 72.8$\pm$0.9& 74.3$\pm$0.5& 72.3$\pm$1.1& 14.4$\pm$6.2& 8.5$\pm$2.5
        &74.5$\pm$0.6& 71.3$\pm$0.4& 72.5$\pm$0.2& 76.9$\pm$1.2& 6.8$\pm$3.1& 4.8$\pm$0.6 \\
        
        \rowcolor{blue!5} \cellcolor{nocolor} &\revise{Avg.} &74.8$\pm$0.3& 68.0$\pm$2.2& 76.5$\pm$3.8& 70.6$\pm$2.0& 75.9$\pm$0.4& 15.5$\pm$1.3& \textbf{9.2}$\pm$2.3
        &74.4$\pm$0.2& 68.9$\pm$1.9& 74.9$\pm$0.3& 71.6$\pm$3.6& 72.3$\pm$2.1& 45.2$\pm$11.0& 29.8$\pm$6.4
        &74.5$\pm$0.3& 72.3$\pm$1.3& 74.8$\pm$0.3& 71.5$\pm$1.8& \textbf{14.4}$\pm$4.8& \textbf{8.5}$\pm$2.5
        &74.5$\pm$0.3& 72.2$\pm$0.5& 73.3$\pm$0.3& 77.0$\pm$0.7& 5.9$\pm$3.2& 3.9$\pm$1.2 \\
        
        \midrule
        \multirow{4}{*}{\shortstack{FairLoRA \\ (Ours)}}
        &C1   &78.2$\pm$0.4& 72.8$\pm$1.6& 82.7$\pm$0.7& 76.3$\pm$1.4& 77.1$\pm$0.3& 14.2$\pm$2.6& 23.1$\pm$2.0&76.2$\pm$0.4& 69.0$\pm$4.9& 76.1$\pm$0.5& 68.0$\pm$6.3& 74.4$\pm$2.7& 50.0$\pm$1.7& 37.3$\pm$5.1&80.1$\pm$0.2& 78.4$\pm$0.9& 80.2$\pm$0.1& 77.9$\pm$1.0& 14.2$\pm$2.2& 14.0$\pm$2.1&78.9$\pm$0.5& 78.0$\pm$0.3& 78.6$\pm$0.4& 79.7$\pm$0.7& 4.3$\pm$0.1& 2.1$\pm$0.1\\ 
        &C2   &80.2$\pm$0.6& 73.0$\pm$0.4& 80.5$\pm$1.9& 72.7$\pm$0.5& 81.2$\pm$0.5& 19.8$\pm$0.5& 18.0$\pm$0.2&78.5$\pm$0.9& 68.0$\pm$0.5& 77.8$\pm$0.9& 87.1$\pm$5.5& 84.8$\pm$1.5& 33.1$\pm$4.3& 29.8$\pm$4.3&77.7$\pm$1.3& 75.8$\pm$1.5& 77.9$\pm$1.3& 75.3$\pm$1.6& 9.4$\pm$1.6& 5.8$\pm$4.9&80.3$\pm$0.1& 78.5$\pm$0.3& 79.2$\pm$0.2& 81.5$\pm$0.0& 7.3$\pm$1.3& 6.0$\pm$2.0\\ 
        &C3   &78.7$\pm$0.2& 70.9$\pm$1.5& 81.9$\pm$1.4& 71.0$\pm$0.7& 78.8$\pm$0.1& 28.0$\pm$3.4& 27.4$\pm$3.2&77.6$\pm$0.9& 65.4$\pm$5.7& 77.3$\pm$1.0& 83.9$\pm$10.3& 86.6$\pm$0.6& 43.9$\pm$13.4& 40.2$\pm$6.1&80.1$\pm$1.4& 75.5$\pm$2.1& 80.0$\pm$1.5& 86.1$\pm$0.5& 19.5$\pm$5.1& 10.0$\pm$3.4&79.3$\pm$0.6& 73.3$\pm$0.1& 74.0$\pm$0.0& 82.2$\pm$0.9& 8.1$\pm$0.9& 2.8$\pm$0.7 \\ 
        \rowcolor{blue!5} \cellcolor{nocolor} &Avg. &\textbf{79.0}$\pm$0.4& \textbf{72.2}$\pm$1.2& \textbf{81.7}$\pm$1.3& 73.3$\pm$0.9& \textbf{79.0}$\pm$0.3& 20.7$\pm$2.2& 22.8$\pm$1.8&\textbf{77.4}$\pm$0.7& \textbf{67.4}$\pm$3.7& \textbf{77.1}$\pm$0.8& 79.7$\pm$7.4& \textbf{81.9}$\pm$1.6& 42.3$\pm$6.5& 35.8$\pm$5.2&\textbf{79.3}$\pm$1.0& \textbf{76.5}$\pm$1.5& \textbf{79.3}$\pm$1.0& \textbf{79.8}$\pm$1.0& \textbf{14.4}$\pm$3.0& 9.9$\pm$3.5&\textbf{79.5}$\pm$0.4& \textbf{76.6}$\pm$0.2& 77.3$\pm$0.2& \textbf{81.1}$\pm$0.6& 6.6$\pm$0.8& 3.6$\pm$0.9 \\
    \bottomrule
    \end{NiceTabular}}
    
    \label{tab:vit_slo}
    \vspace{-2mm}
\end{table*}

%% file: tables/table2_slo_r50.tex
\begin{table*}[!t]
    \centering
    \caption{
    Fairness and performance comparison on \textbf{FairFedMed-Oph (2D SLO Fundus)} images with \textbf{ResNet50} backbone.
    }
    \resizebox{0.99\textwidth}{!}{
    \setlength{\tabcolsep}{1.5pt}
    \begin{NiceTabular}{lc|ccccccc|ccccccc|cccccc|cccccc}
    \toprule
       \multicolumn{2}{c|}{\textbf{Attribute}} &\multicolumn{7}{c|}{\textbf{Race}} &\multicolumn{7}{c|}{\textbf{Language}} &\multicolumn{6}{c}{\textbf{Ethnicity}} &\multicolumn{6}{c}{\revise{\textbf{Gender}}} \\
        \midrule
       Model &\shortstack{Client \\ ID}  &\shortstack{Overall \\ AUC$\uparrow$} &\shortstack{ES- \\ AUC$\uparrow$} &\shortstack{Asian \\ AUC$\uparrow$} &\shortstack{Black \\ AUC$\uparrow$} &\shortstack{White \\ AUC$\uparrow$} &EOD$\downarrow$&SPD$\downarrow$&\shortstack{Overall \\ AUC$\uparrow$} &\shortstack{ES- \\ AUC$\uparrow$} &\shortstack{\!English\! \\ AUC$\uparrow$} &\shortstack{Spanish \\ AUC$\uparrow$} &\shortstack{Others \\ AUC$\uparrow$} &EOD$\downarrow$&SPD$\downarrow$&\shortstack{Overall \\ AUC$\uparrow$} &\shortstack{ES- \\ AUC$\uparrow$} &\shortstack{\!NonHisp.\! \\ AUC$\uparrow$} &\shortstack{Hisp. \\ AUC$\uparrow$} &EOD$\downarrow$&SPD$\downarrow$ &\shortstack{\revise{ Overall} \\ \revise{AUC$\uparrow$}} &\shortstack{\revise{ ES }\\ \revise{ AUC$\uparrow$}} &\shortstack{\revise{ Male} \\ \revise{ AUC$\uparrow$}} &\shortstack{\revise{ Female} \\ \revise{ AUC$\uparrow$}} &\revise{ EOD$\downarrow$} &\revise{ SPD$\downarrow$} \\
        \midrule
       {FedAvg}  &Avg. &73.7$\pm$0.2 &{67.8}$\pm$0.4 &{78.1}$\pm$1.2 &{70.7}$\pm$0.4 &71.5$\pm$0.2 &22.8$\pm$0.8 &20.8$\pm$2.2 &73.5$\pm$0.0 &57.1$\pm$0.4 &73.9$\pm$0.1 &69.2$\pm$0.3 &{81.7}$\pm$0.2 &36.2$\pm$0.0 &34.9$\pm$0.8 &71.8$\pm$0.2 &{66.9}$\pm$0.5 &75.6$\pm$0.2 &69.3$\pm$0.6 & \textbf{17.2}$\pm$0.1 &{\textbf{7.7}}$\pm$0.4 &71.4$\pm$0.3 & 66.4$\pm$0.0 & 72.2$\pm$0.5 & 70.6$\pm$0.1 & 5.8$\pm$0.7 & 4.0$\pm$0.5 \\
        {FedHEAL} &Avg. &{74.1}$\pm$0.1 &66.3$\pm$0.3 &79.2$\pm$0.3 &70.5$\pm$0.3 &70.9$\pm$0.1 &30.0$\pm$1.5 &24.1$\pm$1.3 &{74.5}$\pm$0.3 &{60.0}$\pm$0.3 &{76.2}$\pm$0.0 &70.8$\pm$0.1 &\textbf{82.9}$\pm$0.3 &{35.4}$\pm$1.9 &29.0$\pm$0.4 &{72.7}$\pm$0.1 &65.3 $\pm$0.2 &{76.5}$\pm$0.1 &64.5$\pm$0.1 &{18.4}$\pm$0.7 &8.1$\pm$0.5 & 72.9$\pm$0.1 & 66.8$\pm$0.1 & 74.6$\pm$0.1 & 71.4$\pm$0.1 & 6.8$\pm$1.2 & 4.3$\pm$0.4 \\
        
         PromptFL &Avg. & 70.8$\pm$0.3 & 65.4$\pm$1.0 & 69.2$\pm$1.5 & 67.7$\pm$0.6 & 71.1$\pm$0.3 & \textbf{9.1}$\pm$2.1 & 9.7$\pm$1.0 & 71.5$\pm$0.2 & 64.2$\pm$0.9 & 71.4$\pm$0.2 & 79.5$\pm$1.2 & 70.4$\pm$1.4 & \textbf{20.9}$\pm$3.8 & 14.4$\pm$3.9 &70.5$\pm$0.2 & 62.6$\pm$0.4 & 70.3$\pm$0.2 & \textbf{79.8}$\pm$0.6 & 21.9$\pm$5.8 & 13.5$\pm$3.2 & 72.5$\pm$0.3 & 70.8$\pm$0.6 & 71.7$\pm$0.4 & 73.0$\pm$0.6 & \textbf{4.8}$\pm$1.4 & \textbf{2.6}$\pm$0.8 \\
        FedOTP &Avg. & 70.2$\pm$0.2 & 64.8$\pm$1.1 & 69.1$\pm$1.3 & 67.3$\pm$0.8 & 70.7$\pm$0.3 & 10.9$\pm$2.5 & \textbf{7.7}$\pm$1.1 & 70.4$\pm$0.5 & 64.6$\pm$1.3 & 70.3$\pm$0.5 & 75.4$\pm$2.7 & 70.2$\pm$1.0 & 23.9$\pm$5.2 & \textbf{14.2}$\pm$1.4 & 69.6$\pm$0.2 & 60.5$\pm$1.3 & 69.4$\pm$0.2 & 79.4$\pm$2.2 & 19.7$\pm$1.3 & 11.2$\pm$2.6 & 70.8$\pm$0.4 & 69.5$\pm$0.2 & 70.2$\pm$0.3 & 71.6$\pm$0.6 & 5.0$\pm$2.0 & 3.7$\pm$0.8 \\
         \multirow{1}{*}{FairLoRA} &Avg. &\textbf{78.5}$\pm$0.4 & \textbf{72.9}$\pm$1.3 & \textbf{81.7}$\pm$2.2 & \textbf{74.8}$\pm$1.1 & \textbf{78.3}$\pm$0.5 & 14.8$\pm$5.1 & 17.4$\pm$4.3 &\textbf{78.6}$\pm$0.7 & \textbf{68.9}$\pm$3.8 & \textbf{78.4}$\pm$0.8 & \textbf{81.6}$\pm$7.3 & 78.6$\pm$4.1 & 39.7$\pm$8.5 & 39.6$\pm$5.8 &\textbf{78.1}$\pm$1.2 & \textbf{73.6}$\pm$2.4 & \textbf{78.3}$\pm$1.1 & 75.6$\pm$5.1 & 18.8$\pm$4.4 & 10.2$\pm$2.2 &\textbf{78.4}$\pm$0.5 & \textbf{75.3}$\pm$0.8 & \textbf{76.1}$\pm$0.5 & \textbf{80.3}$\pm$0.8 & 6.1$\pm$1.6 & 4.3$\pm$1.6 \\
    \bottomrule
    \end{NiceTabular}}
    
    \label{tab:r50_slo}
\end{table*}

%% file: tables/table3_oct_vitb.tex
\begin{table*}[!t]
    \centering
     \caption{
     Performance and fairness comparison on \textbf{FairFedMed-Oph (3D OCT B-Scan)} images with \textbf{ViT-B/16} backbone.
     }
    \resizebox{0.99\textwidth}{!}{
    \setlength{\tabcolsep}{1.5pt}
    \begin{NiceTabular}{lc|ccccccc|ccccccc|cccccc|cccccc}
    \toprule
      \multicolumn{2}{c|}{\textbf{Attribute}} &\multicolumn{7}{c|}{\textbf{Race}} &\multicolumn{7}{c|}{\textbf{Language}} &\multicolumn{6}{c}{\textbf{Ethnicity}} &\multicolumn{6}{c}{\revise{\textbf{Gender}}} \\
       \midrule
       Model &\shortstack{Client \\ ID} &\shortstack{Overall \\ AUC$\uparrow$} &\shortstack{ES \\ AUC$\uparrow$} &\shortstack{Asian \\ AUC$\uparrow$} &\shortstack{Black \\ AUC$\uparrow$} &\shortstack{White \\ AUC$\uparrow$} &EOD$\downarrow$ &SPD$\downarrow$ &\shortstack{Overall \\ AUC$\uparrow$} &\shortstack{ES \\ AUC$\uparrow$} &\shortstack{\!English\! \\ AUC$\uparrow$} &\shortstack{Spanish \\ AUC$\uparrow$} &\shortstack{Others \\ AUC$\uparrow$} &EOD$\downarrow$ &SPD$\downarrow$ &\shortstack{Overall \\ AUC$\uparrow$} &\shortstack{ES \\ AUC$\uparrow$} &\shortstack{\!NonHisp.\! \\ AUC$\uparrow$} &\shortstack{Hisp. \\ AUC$\uparrow$} &EOD$\downarrow$ &SPD$\downarrow$ &\shortstack{\revise{ Overall} \\ \revise{AUC$\uparrow$}} &\shortstack{\revise{ ES }\\ \revise{ AUC$\uparrow$}} &\shortstack{\revise{ Male} \\ \revise{ AUC$\uparrow$}} &\shortstack{\revise{ Female} \\ \revise{ AUC$\uparrow$}} &\revise{ EOD$\downarrow$} &\revise{ SPD$\downarrow$} \\
       \midrule
       \multirow{4}{*}{\rotatebox{0}{FedAvg}}
        &C1 &76.4$\pm$0.6 &69.5$\pm$0.3 &71.6$\pm$0.4 &74.2$\pm$0.4 &79.3$\pm$0.8  &34.2$\pm$1.9  &31.8$\pm$2.2 &76.6$\pm$0.5 &60.5$\pm$0.3 &78.1$\pm$0.1 &89.9$\pm$0.4 &88.1$\pm$0.7 &26.5$\pm$4.3 &25.5$\pm$1.3 &79.5$\pm$0.1 & 71.1$\pm$1.0 & 79.2$\pm$0.1 & 91.0$\pm$1.4 & 18.5$\pm$3.4 & 18.8$\pm$2.0 &77.3$\pm$0.0 & 74.4$\pm$0.0 & 78.7$\pm$0.0 & 74.7$\pm$0.0 & 4.0$\pm$0.4 & 3.1$\pm$0.2)\\
        &C2 &75.3$\pm$0.3 &66.6$\pm$0.5 &83.0$\pm$0.9 &71.5$\pm$0.1 &73.8$\pm$0.2 &23.5$\pm$1.3 & 21.8$\pm$0.9 &76.9$\pm$0.2 &68.1$\pm$0.3 &80.3$\pm$1.7 &74.9$\pm$0.5 &69.3$\pm$2.4 & 21.4$\pm$1.0 & 19.6$\pm$1.3 &74.6$\pm$0.2 & 69.9$\pm$0.8 & 74.9$\pm$0.1 & 68.0$\pm$1.1 & 20.0$\pm$1.1 & 15.1$\pm$0.1 &80.2$\pm$0.0 & 80.1$\pm$0.0 & 80.3$\pm$0.0 & 80.3$\pm$0.0 & 5.9$\pm$0.0 & 5.0$\pm$0.1 \\
        &C3 &77.6$\pm$0.5 &69.4$\pm$0.2 &81.2$\pm$0.5 &72.6$\pm$0.1 &74.4$\pm$0.3 & 19.4$\pm$0.7 & 14.6$\pm$0.6 &75.5$\pm$0.1 &65.7$\pm$0.6 &75.3$\pm$0.9 &85.3$\pm$1.1 &80.3$\pm$0.8 & 22.8$\pm$0.2 & 17.3$\pm$0.3 &77.0$\pm$0.0 & 71.8$\pm$0.3 & 76.7$\pm$0.0 & 83.9$\pm$0.4 & 1.6$\pm$0.7 & 2.2$\pm$0.4 &78.6$\pm$0.0 & 74.9$\pm$0.0 & 82.1$\pm$0.0 & 77.1$\pm$0.0 & 9.9$\pm$0.0 & 7.7$\pm$0.1 \\
        \rowcolor{blue!5} \cellcolor{nocolor} &Avg. &76.4$\pm$0.4 &{68.5}$\pm$0.3 &78.6$\pm$0.6 &{72.8}$\pm$0.2 &75.8$\pm$0.4 &25.7$\pm$1.3 &22.7$\pm$1.2 &76.4$\pm$0.3 &64.8$\pm$0.4 &{77.9}$\pm$0.9 &83.3$\pm$0.7 &79.4$\pm$1.3 &24.1$\pm$1.8 &20.9$\pm$1.0 &77.1$\pm$0.1 & 75.7$\pm$0.2 & 77.0$\pm$0.1 & 78.9$\pm$0.5 & 9.6$\pm$0.6 & 11.3$\pm$0.5 &78.8$\pm$0.0 & 77.0$\pm$0.0 & 80.0$\pm$0.0 & 77.7$\pm$0.0 & 5.8$\pm$0.1 & 4.9$\pm$0.0 \\ 
      \midrule
        \multirow{4}{*}{\rotatebox{0}{FedHEAL}}
        &C1 &75.2$\pm$0.1 & 66.1$\pm$0.0 & 86.5$\pm$0.1 & 73.5$\pm$0.2 & 74.4$\pm$0.2 & 21.1$\pm$0.2 & 24.3$\pm$0.2 &76.1$\pm$0.0 & 73.1$\pm$0.9 & 76.2$\pm$0.0 & 75.0$\pm$0.0 & 73.1$\pm$1.3 & 23.2$\pm$0.4 & 25.1$\pm$0.3 &76.1$\pm$0.1 & 65.7$\pm$0.1 & 75.7$\pm$0.0 & 91.4$\pm$0.0 & 20.8$\pm$0.0 & 4.1$\pm$0.1 &73.7$\pm$0.1 & 70.9$\pm$0.2 & 75.0$\pm$0.1 & 71.1$\pm$0.1 & 3.3$\pm$0.6 & 2.3$\pm$1.5 \\
        &C2 &74.0$\pm$0.1 & 71.8$\pm$0.0 & 75.3$\pm$0.2 & 72.2$\pm$0.1 & 74.1$\pm$0.1 & 12.2$\pm$1.3 & 13.4$\pm$0.3 &72.9$\pm$0.1 & 66.8$\pm$0.3 & 73.3$\pm$0.0 & 68.3$\pm$0.6 & 68.9$\pm$0.0 & 12.9$\pm$3.8 & 11.4$\pm$2.9 &72.6$\pm$0.1 & 66.2$\pm$0.2 & 73.0$\pm$0.1 & 63.2$\pm$0.2 & 7.6$\pm$1.3 & 9.6$\pm$0.6 &77.2$\pm$0.1 & 75.5$\pm$0.1 & 78.5$\pm$0.1 & 76.2$\pm$0.1 & 3.1$\pm$1.1 & 0.6$\pm$0.4 \\
        &C3 &76.6$\pm$0.2 & 71.1$\pm$0.2 & 72.3$\pm$0.2 & 80.0$\pm$0.2 & 76.5$\pm$0.2 & 4.7$\pm$0.6 & 14.7$\pm$1.0 &75.4$\pm$0.1 & 54.3$\pm$1.0 & 75.5$\pm$0.0 & 43.2$\pm$2.3 & 82.0$\pm$0.3 & 30.8$\pm$0.0 & 13.6$\pm$0.2 &74.9$\pm$0.0 & 71.5$\pm$0.6 & 74.8$\pm$0.0 & 79.5$\pm$0.9 & 8.0$\pm$0.1 & 6.7$\pm$0.1 &76.4$\pm$0.1 & 72.5$\pm$0.0 & 79.8$\pm$0.2 & 74.4$\pm$0.0 & 11.4$\pm$0.8 & 6.1$\pm$0.3 \\
        \rowcolor{blue!5} \cellcolor{nocolor} &Avg. &75.3$\pm$0.2 & 73.4$\pm$0.0 & 76.8$\pm$0.3 & 74.5$\pm$0.1 & 75.1$\pm$0.1 & \textbf{9.4}$\pm$0.5 & 14.9$\pm$0.4 &74.8$\pm$0.1 & 67.1$\pm$0.2 & 75.0$\pm$0.0 & 64.9$\pm$0.5 & 73.6$\pm$0.2 & \textbf{10.3}$\pm$1.4 & \textbf{5.5}$\pm$1.4 &74.6$\pm$0.1 & 74.2$\pm$0.1 & 74.5$\pm$0.0 & 74.5$\pm$0.5 & \textbf{4.7}$\pm$0.8 & \textbf{4.1}$\pm$0.4 &75.8$\pm$0.1 & 73.5$\pm$0.1 & 77.5$\pm$0.1 & 74.3$\pm$0.1 & 4.0$\pm$0.9 & \textbf{2.0}$\pm$0.6 \\ 
       \midrule
        \multirow{4}{*}{\rotatebox{0}{PromptFL}}
        &C1 &78.3$\pm$0.1& 64.8$\pm$0.5& 64.4$\pm$0.5& 72.5$\pm$0.5& 79.6$\pm$0.1& 23.8$\pm$3.8& 19.0$\pm$1.5&78.0$\pm$0.2& 72.7$\pm$0.8& 78.0$\pm$0.1& 79.0$\pm$2.9& 82.4$\pm$1.4& 25.8$\pm$10.8& 8.9$\pm$3.7&77.4$\pm$0.1& 76.4$\pm$0.7& 77.4$\pm$0.1& 76.2$\pm$0.9& 16.3$\pm$3.0& 13.5$\pm$1.8&78.4$\pm$0.2& 78.2$\pm$0.2& 78.4$\pm$0.2& 78.3$\pm$0.1& 6.5$\pm$1.1& 6.9$\pm$1.2\\ 
        &C2 &78.3$\pm$0.1& 68.9$\pm$0.2& 86.5$\pm$0.2& 73.0$\pm$0.3& 78.3$\pm$0.1& 15.1$\pm$2.8& 13.1$\pm$1.6&77.5$\pm$0.2& 68.2$\pm$1.3& 78.2$\pm$0.1& 72.4$\pm$1.6& 69.6$\pm$0.9& 17.5$\pm$3.3& 10.3$\pm$3.3&75.6$\pm$0.4& 69.5$\pm$0.3& 76.0$\pm$0.4& 67.2$\pm$0.5& 11.9$\pm$3.2& 7.7$\pm$1.1&79.4$\pm$0.2& 77.0$\pm$0.2& 78.0$\pm$0.2& 81.1$\pm$0.2& 6.9$\pm$1.1& 5.6$\pm$0.7\\
        &C3 &78.1$\pm$0.1& 70.4$\pm$0.7& 84.2$\pm$0.6& 73.4$\pm$1.1& 78.3$\pm$0.1& 15.9$\pm$3.3& 6.9$\pm$2.6&76.4$\pm$0.3& 64.2$\pm$1.2& 76.2$\pm$0.4& 87.5$\pm$0.0& 84.3$\pm$1.4& 18.4$\pm$5.3& 15.6$\pm$2.4&79.5$\pm$0.3& 66.2$\pm$0.5& 79.0$\pm$0.3& 99.0$\pm$0.5& 23.8$\pm$3.1& 19.3$\pm$2.2&78.8$\pm$0.2& 74.8$\pm$0.2& 75.4$\pm$0.2& 80.7$\pm$0.2& 9.1$\pm$1.3& 1.1$\pm$0.5\\
        \rowcolor{blue!5} \cellcolor{nocolor} &Avg. &78.3$\pm$0.1& 68.0$\pm$0.5& 78.4$\pm$0.5& 73.0$\pm$0.6& 78.7$\pm$0.1& 18.2$\pm$3.3& 13.0$\pm$1.9&77.3$\pm$0.3& 68.3$\pm$1.1& 77.4$\pm$0.2& 79.6$\pm$1.5& 78.8$\pm$1.2& 20.6$\pm$6.4& 11.6$\pm$3.1&77.5$\pm$0.3& 70.7$\pm$0.5& 77.5$\pm$0.3& 80.8$\pm$0.7& 17.3$\pm$3.1& 13.5$\pm$1.7&78.9$\pm$0.2& 76.7$\pm$0.2& 77.3$\pm$0.2& 80.1$\pm$0.2& 7.5$\pm$1.2& 4.5$\pm$0.8\\
         \midrule
        \multirow{4}{*}{\rotatebox{0}{FedOTP}}
        &C1 &77.8$\pm$0.1& 63.9$\pm$0.3& 63.4$\pm$0.3& 71.9$\pm$0.4& 79.3$\pm$0.0& 24.1$\pm$2.0& 17.9$\pm$1.2&76.4$\pm$0.0& 64.6$\pm$1.8& 76.3$\pm$0.0& 90.1$\pm$4.0& 81.0$\pm$0.8& 36.1$\pm$3.4& 24.0$\pm$2.9&76.6$\pm$0.2& 73.7$\pm$0.7& 76.7$\pm$0.2& 72.8$\pm$0.9& 23.0$\pm$5.9& 18.6$\pm$4.1&77.6$\pm$0.2& 77.3$\pm$0.2& 77.6$\pm$0.1& 77.6$\pm$0.4& 3.9$\pm$1.5& 4.6$\pm$1.7\\
        &C2 &77.6$\pm$0.1& 69.4$\pm$0.2& 84.3$\pm$0.2& 72.5$\pm$0.2& 77.7$\pm$0.1& 13.1$\pm$2.7& 12.7$\pm$1.7&76.4$\pm$0.3& 66.7$\pm$0.8& 77.2$\pm$0.2& 70.7$\pm$1.0& 68.4$\pm$1.0& 16.2$\pm$3.9& 9.5$\pm$2.3&74.4$\pm$0.3& 67.8$\pm$0.4& 74.8$\pm$0.4& 65.2$\pm$0.7& 9.9$\pm$2.8& 5.4$\pm$2.9&79.0$\pm$0.5& 76.4$\pm$0.6& 77.5$\pm$0.6& 80.9$\pm$0.4& 7.5$\pm$1.2& 5.7$\pm$0.5\\
        &C3 &76.5$\pm$0.1& 67.3$\pm$0.2& 82.6$\pm$0.2& 69.4$\pm$0.1& 77.1$\pm$0.1& 19.0$\pm$2.9& 5.6$\pm$4.1&74.8$\pm$0.2& 60.1$\pm$0.3& 74.4$\pm$0.2& 87.5$\pm$0.0& 86.2$\pm$1.0& 27.3$\pm$3.2& 16.9$\pm$2.2&78.4$\pm$0.1& 65.1$\pm$0.4& 77.9$\pm$0.2& 98.3$\pm$0.4& 24.1$\pm$6.4& 16.1$\pm$3.2&78.1$\pm$0.2& 74.6$\pm$0.3& 75.0$\pm$0.3& 79.7$\pm$0.2& 4.2$\pm$2.7& 1.9$\pm$1.6\\
        \rowcolor{blue!5} \cellcolor{nocolor} &Avg. &77.3$\pm$0.1& 66.9$\pm$0.3& 76.8$\pm$0.2& 71.3$\pm$0.2& 78.0$\pm$0.1& 18.8$\pm$2.5& 12.1$\pm$2.3&75.9$\pm$0.2& 63.8$\pm$1.0& 76.0$\pm$0.2& 82.7$\pm$1.7& 78.5$\pm$0.9& 26.5$\pm$3.5& 16.8$\pm$2.5&76.5$\pm$0.2& 68.9$\pm$0.5& 76.5$\pm$0.2& 78.8$\pm$0.7& 19.0$\pm$5.0& 13.4$\pm$3.4&78.2$\pm$0.3& 76.1$\pm$0.4& 76.7$\pm$0.4& 79.4$\pm$0.3& 5.2$\pm$1.8& 4.1$\pm$1.3\\ 
        
        \midrule
        \multirow{4}{*}{\rotatebox{0}{\revise{ViTAdapter}}}
        &\revise{C1} &75.4$\pm$0.1& 68.3$\pm$2.1& 75.3$\pm$2.8& 69.8$\pm$1.6& 74.5$\pm$0.9& 15.3$\pm$3.8& 8.7$\pm$3.2
        &75.4$\pm$0.9& 65.8$\pm$3.6&  73.7$\pm$1.0& 67.7$\pm$10.6& 69.8$\pm$4.8 & 47.4$\pm$17.0& 37.2$\pm$5.1
        &75.5$\pm$0.6& 67.1$\pm$3.1& 73.9$\pm$0.8& 64.3$\pm$5.2& 19.2$\pm$3.1& 9.9$\pm$8.3
        &75.6$\pm$0.2& 70.7$\pm$1.1& 71.8$\pm$1.0& 75.7$\pm$1.2& 2.4$\pm$0.9& 1.2$\pm$0.8\\ 
        
        &\revise{C2} &75.6$\pm$0.9& 68.7$\pm$0.8& 77.5$\pm$0.2& 71.0$\pm$0.7& 75.9$\pm$0.5& 17.7$\pm$2.7& 5.5$\pm$4.0
        &75.3$\pm$0.3& 67.6$\pm$2.7& 74.1$\pm$0.3& 78.4$\pm$7.5& 71.4$\pm$1.4& 30.9$\pm$5.8& 18.3$\pm$7.9
        &76.4$\pm$0.3& 68.7$\pm$3.2& 73.8$\pm$0.7& 74.0$\pm$10.7& 24.7$\pm$0.4& 12.9$\pm$8.5
        & 75.9$\pm$0.5& 72.6$\pm$1.7& 73.3$\pm$1.2& 75.5$\pm$0.8& 5.0$\pm$2.3& 4.2$\pm$1.0 \\
        
        &\revise{C3} &74.5$\pm$0.1& 67.3$\pm$0.2& 76.5$\pm$1.8& 63.4$\pm$2.0& 73.9$\pm$0.7& 18.0$\pm$1.7& 10.5$\pm$2.9
        &74.3$\pm$0.3& 65.1$\pm$4.8& 72.9$\pm$0.7& 67.0$\pm$8.0& 78.5$\pm$0.7& 48.2$\pm$5.8& 39.3$\pm$6.5
        &74.4$\pm$0.4& 70.4$\pm$0.8& 72.6$\pm$0.7& 69.7$\pm$0.9& 14.3$\pm$0.8& 11.1$\pm$4.8
        &74.2$\pm$0.7& 70.2$\pm$1.3& 71.2$\pm$1.1& 75.0$\pm$0.9& 3.7$\pm$1.5& 2.7$\pm$2.2\\
        
        \rowcolor{blue!5} \cellcolor{nocolor} &\revise{Avg.} &75.2$\pm$0.7& 68.1$\pm$1.0& 76.4$\pm$1.6& 68.1$\pm$1.4& 74.8$\pm$0.7& 17.0$\pm$2.7& \textbf{8.2}$\pm$3.3
        &75.0$\pm$0.6& 66.2$\pm$3.7& 73.6$\pm$0.7& 71.1$\pm$8.7& 73.2$\pm$2.3& 42.2$\pm$9.5& 31.6$\pm$6.5
        &75.4$\pm$0.4& 68.7$\pm$2.4& 73.5$\pm$0.8& 69.3$\pm$5.6& 19.4$\pm$1.4& 11.3$\pm$7.2
        &75.2$\pm$0.4& 71.2$\pm$ 1.4& 72.1$\pm$1.1& 75.4$\pm$1.0& \textbf{3.7}$\pm$1.6& 2.7$\pm$1.3 \\
        
        \midrule
        \multirow{4}{*}{\rotatebox{0}{\shortstack{FairLoRA \\ (Ours)}}}
        &C1 &83.7$\pm$0.4& 73.1$\pm$0.3& 73.6$\pm$1.3& 79.7$\pm$0.8& 84.1$\pm$0.5& 19.7$\pm$3.7& 24.9$\pm$1.3&81.9$\pm$0.3& 72.1$\pm$0.6& 81.6$\pm$0.3& 90.3$\pm$1.0& 86.9$\pm$0.8& 33.2$\pm$3.6& 34.1$\pm$2.8&81.9$\pm$1.4& 80.9$\pm$2.2& 82.0$\pm$1.3& 82.2$\pm$1.6& 15.7$\pm$8.6& 14.7$\pm$6.5&84.2$\pm$0.7& 81.9$\pm$0.8& 83.3$\pm$0.7& 86.2$\pm$0.6& 4.7$\pm$2.6& 5.0$\pm$1.3\\
        &C2 &82.8$\pm$0.3& 72.9$\pm$0.5& 91.3$\pm$0.5& 78.3$\pm$0.2& 82.2$\pm$0.4& 17.2$\pm$3.7& 16.0$\pm$0.8&82.5$\pm$0.3& 78.8$\pm$1.6& 82.5$\pm$0.3& 84.1$\pm$0.6& 80.1$\pm$2.4& 27.6$\pm$9.1& 27.2$\pm$6.0&81.5$\pm$0.2& 77.2$\pm$2.0& 81.8$\pm$0.3& 76.0$\pm$2.7& 10.5$\pm$4.7& 9.9$\pm$6.9&85.0$\pm$0.2& 82.8$\pm$0.6& 83.8$\pm$0.4& 86.4$\pm$0.1& 5.8$\pm$1.1& 4.3$\pm$1.3\\
        &C3 &83.4$\pm$0.4& 76.1$\pm$0.7& 88.3$\pm$0.1& 78.8$\pm$0.6& 83.2$\pm$0.3& 16.2$\pm$3.3& 11.9$\pm$2.7&82.6$\pm$0.3& 76.2$\pm$2.6& 82.5$\pm$0.3& 87.5$\pm$0.0& 79.0$\pm$3.6& 39.4$\pm$14.6& 28.9$\pm$7.3&83.6$\pm$0.4& 75.4$\pm$0.6& 83.4$\pm$0.4& 94.3$\pm$1.7& 16.2$\pm$4.4& 7.6$\pm$4.1&83.2$\pm$0.5& 79.3$\pm$0.8& 79.9$\pm$0.8& 84.8$\pm$0.5& 7.5$\pm$1.1& 3.4$\pm$1.6\\
        \rowcolor{blue!5} \cellcolor{nocolor} &Avg. &\textbf{83.3}$\pm$\textbf{}0.3& \textbf{74.0}$\pm$0.5& \textbf{84.4}$\pm$0.6& \textbf{78.9}$\pm$0.5& \textbf{83.2}$\pm$0.4& 17.7$\pm$3.5& 17.6$\pm$1.6&\textbf{82.3}$\pm$0.3& \textbf{75.7}$\pm$1.6& \textbf{82.2}$\pm$0.3& \textbf{87.3}$\pm$0.5& \textbf{82.0}$\pm$2.2& 33.4$\pm$9.1& 30.1$\pm$5.4&\textbf{82.4}$\pm$0.7& \textbf{77.8}$\pm$1.6& \textbf{82.4}$\pm$0.7& \textbf{84.2}$\pm$2.0& 14.1$\pm$5.9& 10.7$\pm$5.8&\textbf{84.1}$\pm$0.5& \textbf{81.3}$\pm$0.7& \textbf{82.3}$\pm$0.6& \textbf{85.8}$\pm$0.4& 6.0$\pm$1.6& 4.3$\pm$1.4\\
    \bottomrule
    \end{NiceTabular}}
   
    \label{tab:vit_oct}
\end{table*}

%% file: tables/table4_oct_r50.tex
\begin{table*}[!t]
    \centering
    \caption{
    Performance and fairness comparison on \textbf{FairFedMed-Oph (3D OCT B-Scan)} images with \textbf{ResNet50} backbone.
    }
    \resizebox{0.99\textwidth}{!}{
    \setlength{\tabcolsep}{1.5pt}
    \begin{NiceTabular}{lc|ccccccc|ccccccc|cccccc|cccccc}
    \toprule
       \multicolumn{2}{c|}{\textbf{Attribute}} &\multicolumn{7}{c|}{\textbf{Race}} &\multicolumn{7}{c|}{\textbf{Language}} &\multicolumn{6}{c}{\textbf{Ethnicity}} &\multicolumn{6}{c}{\revise{\textbf{Gender}}} \\
       \midrule
       Model &\shortstack{Client \\ ID}&\shortstack{Overall \\ AUC$\uparrow$} &\shortstack{ES \\ AUC$\uparrow$} &\shortstack{Asian \\ AUC$\uparrow$} &\shortstack{Black \\ AUC$\uparrow$} &\shortstack{White \\ AUC$\uparrow$} &EOD$\downarrow$ &SPD$\downarrow$ &\shortstack{Overall \\ AUC$\uparrow$} &\shortstack{ES \\ AUC$\uparrow$} &\shortstack{\!English\! \\ AUC$\uparrow$} &\shortstack{\!Spanish\! \\ AUC$\uparrow$} &\shortstack{Others \\ AUC$\uparrow$} &EOD$\downarrow$ &SPD$\downarrow$ &\shortstack{Overall \\ AUC$\uparrow$} &\shortstack{ES \\ AUC$\uparrow$} &\shortstack{NonHisp. \\ AUC$\uparrow$} &\shortstack{Hisp. \\ AUC$\uparrow$} &EOD$\downarrow$ &SPD$\downarrow$ &\shortstack{\revise{ Overall} \\ \revise{AUC$\uparrow$}} &\shortstack{\revise{ ES }\\ \revise{ AUC$\uparrow$}} &\shortstack{\revise{ Male} \\ \revise{ AUC$\uparrow$}} &\shortstack{\revise{ Female} \\ \revise{ AUC$\uparrow$}} &\revise{ EOD$\downarrow$} &\revise{ SPD$\downarrow$} \\
       \midrule
      {{FedAvg}}
        &Avg. &76.5$\pm$0.7 & 69.4$\pm$1.4 & 83.3$\pm$0.9 & 73.2$\pm$0.1 & 76.3$\pm$0.9  &25.7$\pm$0.7 &22.7$\pm$0.2 &74.9$\pm$0.0 & 73.6$\pm$0.1 & 74.9$\pm$0.0 & 73.4$\pm$0.1 & 74.9$\pm$0.3 &35.6$\pm$1.8 &32.9$\pm$1.7 &75.7$\pm$0.2 & 71.8$\pm$1.5 & 75.4$\pm$0.2 & 80.9$\pm$1.8 & 21.5$\pm$0.7 & 17.9$\pm$0.3 & 78.2$\pm$0.4 & 75.1$\pm$0.2 & 80.3$\pm$0.4 & 76.2$\pm$0.3 & 8.5$\pm$1.1 & 6.0$\pm$0.7 \\
       {{FedHEAL}}
        &Avg. &75.9$\pm$0.4 & 72.3$\pm$0.4 & 80.1$\pm$1.0 & 75.3$\pm$0.1 & 75.7$\pm$0.2 &24.3$\pm$0.4 &20.6$\pm$0.5 &76.6$\pm$0.5 & 71.8$\pm$0.2 & 76.6$\pm$0.1 & 81.2$\pm$1.0 & 75.6$\pm$0.7 & 26.4$\pm$0.7 & 21.5 $\pm$0.4 &76.3$\pm$0.1 & 70.6$\pm$0.1 & 76.5$\pm$0.1 & 68.5$\pm$0.1 & 18.6$\pm$0.7 & 15.2$\pm$0.5 &78.6$\pm$0.1 & 76.8$\pm$0.2 & 79.8$\pm$0.1 & 77.5$\pm$0.2 & 19.7$\pm$0.7 & 18.0$\pm$0.0 \\
       {{PromptFL}}
        &Avg. &73.0$\pm$0.3 & 66.8$\pm$1.2 & 78.4$\pm$1.6 & 74.0$\pm$1.1 & 72.8$\pm$0.3 & 13.4$\pm$2.5 & 6.0$\pm$2.1 &72.7$\pm$0.4 & 65.4$\pm$2.5 & 72.7$\pm$0.4 & 74.5$\pm$2.4 & 71.7$\pm$2.0 & \textbf{18.7}$\pm$6.0 & \textbf{10.9}$\pm$3.5 &73.1$\pm$0.4 & 69.5$\pm$0.8 & 73.0$\pm$0.5 & 77.0$\pm$1.3 & \textbf{6.5}$\pm$3.6 & \textbf{3.7}$\pm$2.7 &73.8$\pm$0.2 & 71.5$\pm$0.5 & 71.8$\pm$0.5 & 74.7$\pm$0.4 & 6.6$\pm$2.4 & 5.2$\pm$1.9 \\
       {{FedOTP}}
        &Avg. &73.7$\pm$0.2 & 68.2$\pm$0.9 & 79.3$\pm$0.6 & 73.5$\pm$0.8 & 73.6$\pm$0.1 & 13.8$\pm$1.3 & 6.8$\pm$1.2 &73.1$\pm$0.0 & 68.6$\pm$0.0 & 73.2$\pm$0.0 & 72.0$\pm$0.0 & 74.7$\pm$0.0 & 27.7$\pm$0.0 & 11.5$\pm$0.0 &74.1$\pm$0.1 & 67.5$\pm$0.6 & 73.8$\pm$0.1 & 82.3$\pm$0.8 & 13.8$\pm$3.4 & 9.1$\pm$3.2 &74.1$\pm$0.0 & 71.9$\pm$0.0 & 72.3$\pm$0.0 & 74.9$\pm$0.0 & \textbf{5.9}$\pm$0.0 & \textbf{4.7}$\pm$0.1 \\
       \multirow{1}{*}{FairLoRA}
        &Avg. &\textbf{82.9}$\pm$0.9 & \textbf{74.2}$\pm$1.4 & \textbf{85.9}$\pm$1.8 & \textbf{79.4}$\pm$1.4 & \textbf{82.7}$\pm$0.9 & 14.8$\pm$2.7 & 18.1$\pm$2.0 &\textbf{82.6}$\pm$0.3 & \textbf{76.4}$\pm$1.9 & \textbf{82.6}$\pm$0.3 & \textbf{83.9}$\pm$3.1 & \textbf{78.6}$\pm$2.2 & 38.7$\pm$11.0 & 32.6$\pm$7.2 &\textbf{82.6}$\pm$0.5 & \textbf{78.8}$\pm$1.9 & \textbf{82.7}$\pm$0.4 & \textbf{84.0}$\pm$2.8 & 17.0$\pm$8.8 & 9.0$\pm$4.0 &\textbf{84.7}$\pm$0.5 & \textbf{81.8}$\pm$0.7 & \textbf{82.9}$\pm$0.6 & \textbf{86.5}$\pm$0.5 & 9.1$\pm$1.4 & 6.5$\pm$1.3 \\
    \bottomrule
    \end{NiceTabular}}
    
    \label{tab:r50_oct}
\end{table*}

%% file: tables/table5_vit_chexmimic.tex
\begin{table*}[!t]
    \color{black}
    \centering
     \caption{\revise{
     Performance and fairness comparison on \textbf{FairFedMed-Chest} with \textbf{ViT-B/16} backbone. 
     }}\label{tab:vit_chest}
    \resizebox{0.99\textwidth}{!}{
    \setlength{\tabcolsep}{1.5pt}
    \begin{NiceTabular}{lc|ccccccc|cccccc|cccccc}
      \toprule
      \multicolumn{2}{c|}{\textbf{Attribute}} &\multicolumn{7}{c|}{\textbf{Race}} &\multicolumn{6}{c}{\textbf{Gender}} &\multicolumn{6}{c}{\textbf{Age}} \\
       \midrule
       Model &\shortstack{Client \\ ID} &\shortstack{Overall \\ AUC$\uparrow$} &\shortstack{ES \\ AUC$\uparrow$} &\shortstack{Asian \\ AUC$\uparrow$} &\shortstack{Black \\ AUC$\uparrow$} &\shortstack{White \\ AUC$\uparrow$} &EOD$\downarrow$ &SPD$\downarrow$ &\shortstack{Overall \\ AUC$\uparrow$} &\shortstack{ES \\ AUC$\uparrow$} &\shortstack{Male \\ AUC$\uparrow$} &\shortstack{Female \\ AUC$\uparrow$} &EOD$\downarrow$ &SPD$\downarrow$ &\shortstack{Overall \\ AUC$\uparrow$} &\shortstack{ES \\ AUC$\uparrow$} &\shortstack{$0-60$ \\ AUC$\uparrow$} &\shortstack{$60+$ \\ AUC$\uparrow$} &EOD$\downarrow$ &SPD$\downarrow$ \\
       \midrule
        \multirow{4}{*}{\rotatebox{0}{PromptFL}}
        &C1-Chex &72.2$\pm$0.2 & 68.2$\pm$0.7 & 71.4$\pm$0.1 & 74.1$\pm$0.6 & 75.3$\pm$1.2 & 16.2$\pm$2.3 & 11.1$\pm$1.6 &71.0$\pm$0.2 & 69.8$\pm$0.3 & 70.3$\pm$0.2 & 72.0$\pm$0.5 & 1.2$\pm$0.7 & 0.7$\pm$0.6 &71.7$\pm$0.2 & 70.3$\pm$0.2 & 72.8$\pm$0.1 & 70.8$\pm$0.2 & 14.9$\pm$2.2 & 14.5$\pm$0.9 \\
        &C2-Mimic &78.7$\pm$0.2 & 70.8$\pm$0.5 & 78.4$\pm$0.3 & 70.9$\pm$0.6 & 81.7$\pm$1.1 & 21.1$\pm$3.5 & 8.4$\pm$1.3 &78.5$\pm$0.5 & 75.4$\pm$0.5 & 76.6$\pm$0.5 & 80.7$\pm$0.6 & 6.2$\pm$1.4 & 3.1$\pm$1.3 &77.2$\pm$0.1 & 73.3$\pm$0.2 & 80.8$\pm$0.1 & 75.4$\pm$0.2 & 9.8$\pm$1.6 & 6.9$\pm$0.5 \\
        \rowcolor{blue!5} \cellcolor{nocolor} &Avg. &76.6$\pm$0.1 & 70.0$\pm$0.7 & 76.1$\pm$0.2 & 72.1$\pm$0.6 & 79.6$\pm$1.1 & 19.7$\pm$3.2 & 9.6$\pm$1.3 &76.0$\pm$0.4 & 73.6$\pm$0.4 & 74.5$\pm$0.4 & 77.8$\pm$0.5 & 4.7$\pm$1.1 & 2.5$\pm$1.0 &75.4$\pm$0.2 & 72.3$\pm$0.3 & 78.2$\pm$0.2 & 73.8$\pm$0.2 & 11.2$\pm$1.7 & 9.4$\pm$0.6 \\ 
         \midrule
        \multirow{4}{*}{\rotatebox{0}{FedOTP}}
        &C1-Chex &73.1$\pm$0.2 & 69.4$\pm$0.3 & 72.4$\pm$0.2 & 74.9$\pm$0.3 & 76.0$\pm$0.2 & 15.4$\pm$5.1 & 11.9$\pm$1.8 &71.1$\pm$0.1 & 68.9$\pm$0.0 & 69.8$\pm$0.0 & 73.0$\pm$0.1 & 4.4$\pm$1.1 & 1.8$\pm$0.8 &72.1$\pm$0.3 & 71.7$\pm$0.4 & 72.2$\pm$0.3 & 71.7$\pm$0.4 & 14.0$\pm$1.7 & 12.6$\pm$2.3 \\
        &C2-Mimic &78.3$\pm$0.1 & 70.3$\pm$0.6 & 77.8$\pm$0.1 & 70.7$\pm$0.8 & 81.5$\pm$0.2 & 18.1$\pm$1.9 & 9.9$\pm$0.9 &78.0$\pm$0.1 & 74.9$\pm$0.2 & 76.1$\pm$0.1 & 80.3$\pm$0.1 & 5.6$\pm$0.6 & 2.2$\pm$0.8 &76.3$\pm$0.2 & 72.4$\pm$0.1 & 79.8$\pm$0.2 & 74.4$\pm$0.2 & 6.3$\pm$0.6 & 5.9$\pm$0.8 \\
        \rowcolor{blue!5} \cellcolor{nocolor} &Avg. &76.6$\pm$0.2 & 70.1$\pm$0.4 & 76.1$\pm$0.2 & 72.2$\pm$0.6 & 79.8$\pm$0.3 & 18.0$\pm$3.0 & 11.1$\pm$1.6 &75.9$\pm$0.1 & 73.1$\pm$0.1 & 74.2$\pm$0.1 & 78.1$\pm$0.1 & 5.0$\pm$0.7 & 2.1$\pm$0.8 &74.9$\pm$0.2 & 72.1$\pm$0.2 & 77.3$\pm$0.2 & 73.5$\pm$0.2 & 9.3$\pm$0.9 & 8.2$\pm$1.2 \\ 

        \midrule
        \multirow{4}{*}{\rotatebox{0}{ViTAdapter}}
        &C1-Chex &78.9$\pm$0.0 & 75.0$\pm$0.5 & 78.2$\pm$0.1 & 80.6$\pm$0.3 & 81.3$\pm$0.3 & 6.1$\pm$2.6 & 8.2$\pm$3.4 
        &78.0$\pm$0.1 & 76.6$\pm$0.7 & 80.1$\pm$0.3 & 77.2$\pm$0.6 & 9.6$\pm$3.0 & 8.4$\pm$0.7
        &79.1$\pm$0.1 & 76.5$\pm$0.2 & 80.6$\pm$0.2 & 77.4$\pm$0.1 & 8.7$\pm$1.2 & 9.6$\pm$0.1 \\
        &C2-Mimic &80.0$\pm$0.6 & 70.0$\pm$0.6 & 79.7$\pm$0.6 & 69.3$\pm$0.7 & 83.3$\pm$1.0 & 12.7$\pm$0.6 & 5.2$\pm$0.1 
        &79.7$\pm$0.2 & 79.7$\pm$0.2 & 80.3$\pm$0.4 & 79.7$\pm$1.0 & 2.3$\pm$1.7 & 2.5$\pm$0.4 
        &80.0$\pm$0.2 & 77.5$\pm$1.0 & 81.5$\pm$0.3 & 78.3$\pm$0.8  & 19.0$\pm$4.6 & 23.6$\pm$2.0 \\
        \rowcolor{blue!5} \cellcolor{nocolor} &Avg. &79.4$\pm$0.8 & 72.5$\pm$3.5 & 78.9$\pm$1.0 & 75.1$\pm$8.2 & 82.3$\pm$1.4 & 9.4$\pm$4.6 & 6.7$\pm$2.1 
        &78.9$\pm$1.2 & 78.1$\pm$2.1 & 80.2$\pm$0.1 & 78.4$\pm$1.8 & 5.9$\pm$5.2 & 5.4$\pm$4.2 
        &79.5$\pm$0.7 & 77.0$\pm$0.7 & 81.0$\pm$0.6 & 77.8$\pm$0.6 & 13.8$\pm$0.7 & 16.6$\pm$0.9 \\ 
        
        \midrule
        \multirow{4}{*}{\rotatebox{0}{\shortstack{FairLoRA \\ (Ours)}}}
        &C1-Chex &81.0$\pm$0.7 & 75.9$\pm$1.4 & 80.5$\pm$1.0 & 80.8$\pm$1.7 & 85.5$\pm$0.5 & 8.8$\pm$3.5 & 8.2$\pm$3.9 &80.0$\pm$0.9 & 78.6$\pm$0.7 & 79.3$\pm$0.7 & 81.1$\pm$1.4 & 3.3$\pm$2.3 & 2.2$\pm$1.7 &81.1$\pm$0.5 & 79.0$\pm$0.9 & 82.7$\pm$0.4 & 80.0$\pm$0.8 & 4.2$\pm$2.8 & 4.5$\pm$3.3 \\
        &C2-Mimic &84.2$\pm$0.6 & 81.2$\pm$1.4 & 83.9$\pm$0.5 & 82.3$\pm$1.9 & 85.6$\pm$1.1 & 6.4$\pm$3.2 & 3.6$\pm$2.6 &83.1$\pm$0.9 & 82.0$\pm$1.1 & 82.5$\pm$1.0 & 83.8$\pm$0.9 & 2.7$\pm$2.2 & 1.3$\pm$1.0 &83.3$\pm$0.8 & 79.6$\pm$0.7 & 86.3$\pm$0.8 & 81.6$\pm$0.8 & 7.8$\pm$1.2 & 1.6$\pm$0.8 \\
        \rowcolor{blue!5} \cellcolor{nocolor} &Avg. &\textbf{82.6}$\pm$0.6 & \textbf{78.6}$\pm$1.4 & \textbf{82.2}$\pm$0.7 & \textbf{81.5}$\pm$1.8 & \textbf{85.6}$\pm$0.8 & \textbf{7.6}$\pm$3.3 & \textbf{5.9}$\pm$3.3 &\textbf{81.6}$\pm$0.9 & \textbf{80.3}$\pm$0.9 & \textbf{80.9}$\pm$0.8 & \textbf{82.5}$\pm$1.1 & \textbf{3.0}$\pm$2.2 & \textbf{1.7}$\pm$1.4 &\textbf{82.2}$\pm$0.7 & \textbf{79.3}$\pm$0.8 & \textbf{84.5}$\pm$0.6 & \textbf{80.8}$\pm$0.8 & \textbf{6.0}$\pm$2.0 & \textbf{3.0}$\pm$2.0 \\
    \bottomrule
    \end{NiceTabular}
    }
\end{table*}

%% file: tables/table6_r50_chexmimic.tex
\begin{table*}[!t]
    \color{black}
    \centering
    \caption{\revise{
    Performance and fairness comparison on \textbf{FairFedMed-Chest} with \textbf{ResNet50} backbone. 
    }}\label{tab:r50_chest}
    \resizebox{0.99\textwidth}{!}{
    \setlength{\tabcolsep}{1.5pt}
    \begin{NiceTabular}{lc|ccccccc|cccccc|cccccc}
    \toprule
       \multicolumn{2}{c|}{\textbf{Attribute}} &\multicolumn{7}{c|}{\textbf{Race}} &\multicolumn{6}{c}{\textbf{Gender}} &\multicolumn{6}{c}{\textbf{Age}} \\
       \midrule
       Model &\shortstack{Client \\ ID}&\shortstack{Overall \\ AUC$\uparrow$} &\shortstack{ES \\ AUC$\uparrow$} &\shortstack{Asian \\ AUC$\uparrow$} &\shortstack{Black \\ AUC$\uparrow$} &\shortstack{White \\ AUC$\uparrow$} &EOD$\downarrow$ &SPD$\downarrow$ &\shortstack{Overall \\ AUC$\uparrow$} &\shortstack{ES \\ AUC$\uparrow$} &\shortstack{Male \\ AUC$\uparrow$} &\shortstack{Female \\ AUC$\uparrow$} &EOD$\downarrow$ &SPD$\downarrow$ &\shortstack{Overall \\ AUC$\uparrow$} &\shortstack{ES \\ AUC$\uparrow$} &\shortstack{$0-60$ \\ AUC$\uparrow$} &\shortstack{$60+$ \\ AUC$\uparrow$} &EOD$\downarrow$ &SPD$\downarrow$ \\
       \midrule
       {{PromptFL}}
        &\quad Avg. \quad&72.9$\pm$0.3  & 70.9$\pm$0.9  & 72.6$\pm$0.3  & 73.0$\pm$1.0  & 74.7$\pm$0.5  & 7.2$\pm$2.8  & 6.0$\pm$2.0  &73.5$\pm$0.2  & 72.3$\pm$0.4  & 73.3$\pm$0.3  & 73.8$\pm$0.4  & 3.9$\pm$1.6  & 2.8$\pm$1.4  &70.9$\pm$0.2  & 68.2$\pm$0.4  & 73.3$\pm$0.5  & 69.5$\pm$0.3  & 11.2$\pm$2.1  & 9.3$\pm$1.6   \\
       {{FedOTP}}
        &\quad Avg. \quad&72.9$\pm$0.2  & 71.1$\pm$0.6  & 72.6$\pm$0.3  & 74.0$\pm$0.6  & 73.8$\pm$0.6  & 8.5$\pm$3.8  & \textbf{5.7}$\pm$1.7  &73.1$\pm$0.1  & 71.8$\pm$0.3  & 72.5$\pm$0.2  & 73.9$\pm$0.3  & \textbf{3.6}$\pm$1.2  & \textbf{1.9}$\pm$0.7  &71.7$\pm$0.1  & 69.0$\pm$0.2  & 74.1$\pm$0.2  & 70.2$\pm$0.2  & 9.8$\pm$1.2  & 8.9$\pm$0.9  \\
       \multirow{1}{*}{FairLoRA}
        &\quad Avg. \quad&\textbf{84.1}$\pm$0.9  & \textbf{81.4}$\pm$1.0  & \textbf{84.0}$\pm$1.0  & \textbf{83.2}$\pm$1.2  & \textbf{85.5}$\pm$1.7  & \textbf{6.7}$\pm$2.7  & 6.7$\pm$2.7  &\textbf{83.9}$\pm$0.6  & \textbf{81.6}$\pm$1.0  & \textbf{82.7}$\pm$0.8  & \textbf{85.5}$\pm$0.6  & 6.9$\pm$3.9  & 3.0$\pm$2.1  &\textbf{83.8}$\pm$0.6  & \textbf{80.9}$\pm$0.5  & \textbf{86.0}$\pm$1.0  & \textbf{82.4}$\pm$0.4  & \textbf{6.2}$\pm$2.3  & \textbf{3.5}$\pm$2.8  \\
    \bottomrule
    \end{NiceTabular}
    }
\end{table*}

%% file: figs/abl_study.tex
\begin{figure*}
    \centering
    \begin{subfigure}{0.72\textwidth}
        \centering
        \includegraphics[width=1\linewidth]{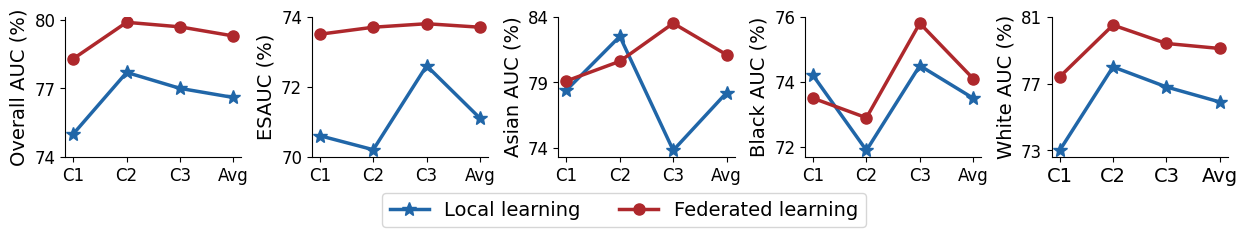}
        \vspace{-6mm}
        \caption{FairLoRA with local learning only and federated learning}\label{fig:single_federated}
    \end{subfigure}
    \begin{subfigure}{0.27\textwidth}
        \centering
        \includegraphics[width=0.9\linewidth]{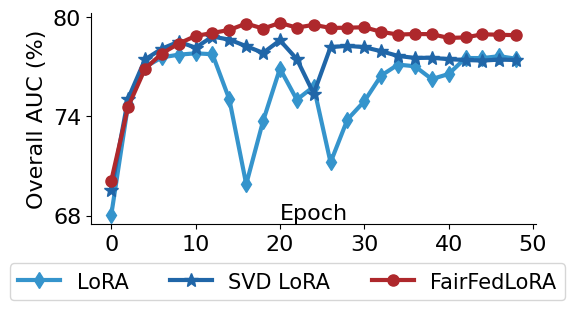}
        \vspace{-1mm}
        \caption{Training convergence of LoRAs}\label{fig:lora_convergence}
    \end{subfigure}%
    \vfill
    \begin{subfigure}{0.72\textwidth}
        \centering
        \includegraphics[width=1\linewidth]{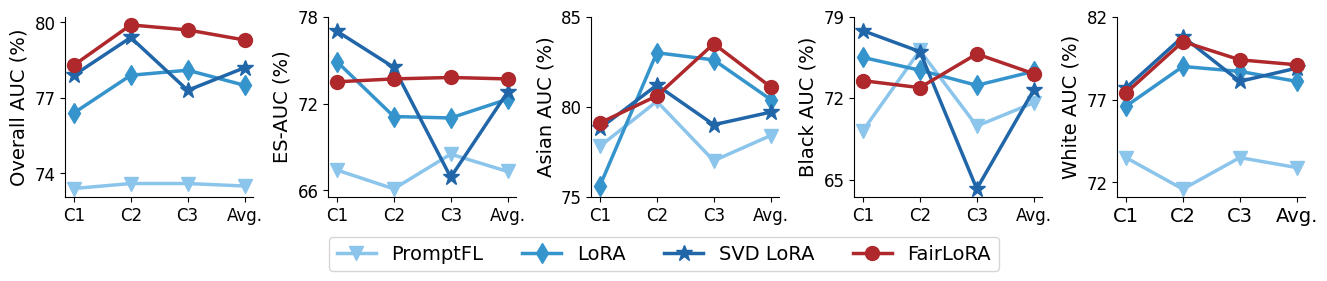}
        \vspace{-6mm}
        \caption{Baseline model PromptFL and three LoRA variants: LoRA, SVD-based LoRA and FairLoRA}\label{fig:LoRAS}
    \end{subfigure}%
    \begin{subfigure}{0.27\textwidth}
        \centering
        \includegraphics[width=0.85\linewidth]{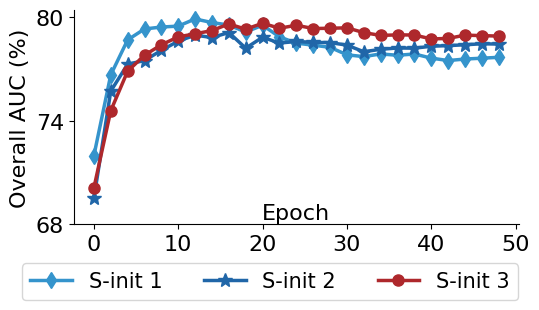}
        \vspace{-2mm}
        \caption{{\footnotesize $\{\overline{\mS}_{g}^0\}_{g\in \mathcal{G}}$} initialization}\label{fig:s_init_abl}
    \end{subfigure}%
    \caption{Ablation study conducted on the \textbf{Race} attribute of \textbf{2D SLO Fundus} images, using the \textbf{ViT-B/16} backbone.}
\end{figure*}

%% file: tables/table8_2_out_3.tex
\begin{table}[!t]
    \centering
    \caption{\revise{Impact of client sampling during local training (ViT-B backbone, ethnicity attribute, 3D OCT B-Scan images in FairFedMed-Oph).}}
    \label{tab:num_clients_local_training}
    \revise{
    \setlength{\tabcolsep}{1.2pt}
    \begin{NiceTabular}{c|cccccc}
         \toprule
         \shortstack{Num. \\ Clients} &\shortstack{Overall \\ AUC$\uparrow$} &\shortstack{ES \\ AUC$\uparrow$} &\shortstack{NonHisp. \\ AUC$\uparrow$} &\shortstack{Hisp. \\ AUC$\uparrow$} &EOD$\downarrow$ &SPD$\downarrow$ \\
         \midrule
         3/3 &\textbf{82.7}$\pm$1.0 &\textbf{79.6}$\pm$2.0 &\textbf{82.8}$\pm$1.7 &\textbf{86.2}$\pm$2.5 &15.7$\pm$4.7 &\textbf{7.2}$\pm$4.2 \\
         2/3 &{82.4}$\pm$0.7& {77.8}$\pm$1.6& {82.4}$\pm$0.7& {84.2}$\pm$2.0 &\textbf{14.1}$\pm$5.9& 10.7$\pm$5.8 \\
         \midrule
         $\Delta$ &$-$0.3 &$-$1.8 &$-$0.4 &$-$2.0 &$-$1.6 &$+$3.5 \\
         \bottomrule
    \end{NiceTabular}
    }
\end{table}

%% file: tables/table7_abl.tex
\begin{table}[!t]
    \centering
    \caption{\revise{FairLoRA performance with and without taking demographic metadata as input during inference (ViT-B, race attribute of 3D OCT B-Scan in FairFedMed-Oph).}}
    \label{tab:wo_demo_inf}
    \revise{
    \setlength{\tabcolsep}{1.2pt}
    \begin{NiceTabular}{c|ccccccc}
        \toprule
         \shortstack{Demo. \\ Group}&\shortstack{Overall \\ AUC$\uparrow$} &\shortstack{ES \\ AUC$\uparrow$} &\shortstack{Asian \\ AUC$\uparrow$} &\shortstack{Black \\ AUC$\uparrow$} &\shortstack{White \\ AUC$\uparrow$} &EOD$\downarrow$ &SPD$\downarrow$ \\
         \midrule
         \cmark &\textbf{83.3}$\pm$0.3 &\textbf{74.0}$\pm$0.5 &\textbf{84.4}$\pm$0.6 &\textbf{78.9}$\pm$0.5 & {83.2}$\pm$0.4 &\textbf{17.7}$\pm$3.5 & 17.6$\pm$1.6 \\
         \xmark &82.8$\pm$0.4 &72.3$\pm$0.6 &83.7$\pm$0.6 &78.7$\pm$0.8 &\textbf{83.3}$\pm$0.3 &18.4$\pm$5.1 &\textbf{15.0}$\pm$3.2 \\
         \midrule
         $\Delta$ &$-$0.5 &$-$1.7 &$-$0.7 &$-$0.2 &$+$0.1 &$+$0.7 &$-$2.6 \\
         \bottomrule
    \end{NiceTabular}
    }
\end{table}

%% file: secs/6_conclusion.tex
\section{Conclusion, Limitations and Future Work}

\revise{
This paper establishes the first comprehensive benchmark for fairness in medical FL. In addition, we introduce FairFedMed, the first group fairness-aware medical FL dataset featuring real-world ophthalmology data for simulated FL and chest X-rays for real cross-site FL. We also propose FairLoRA, a novel fairness-aware low-rank approximation framework that preserves both intra- and inter-group characteristics while enabling global knowledge sharing. Experiments show that FairLoRA achieves strong classification performance and enhanced fairness across demographic groups. By addressing fairness at the group level, this work marks an important step toward equitable and trustworthy FL systems in healthcare.
}

\revise{
\textbf{Limitations and Future Work.} While FairFedMed serves as a valuable dataset for studying group fairness in federated medical imaging, it lacks population diversity. It mainly reflects patients from an affluent urban area, underrepresented groups like Asian, Hispanic, and non-English-speaking individuals. This limits its generalizability for fairness evaluations. Future work will expand FairFedMed with data from more diverse institutions to support more realistic and equitable assessments. Additionally, we plan to implement \textit{FairLoRA} in real-world federated learning deployments to validate its practical effectiveness and robustness. We also aim to extend the framework to support multiple sensitive attributes, facilitating fairness modeling across intersecting dimensions such as race, language, and ethnicity.
}